\RequirePackage{booktabs}
\documentclass[sn-basic,iicol,pdflatex]{sn-jnl}

\usepackage{graphicx}%
\usepackage{multirow}%
\usepackage{amsmath,amssymb,amsfonts}%
\usepackage{amsthm}%
\usepackage{mathrsfs}%
\usepackage[title]{appendix}%
\usepackage{xcolor}%
\usepackage{textcomp}%
\usepackage{array}
\usepackage{multirow}
\usepackage{manyfoot}%
\usepackage{booktabs}%
\usepackage{algorithm}%
\usepackage{algorithmicx}%
\usepackage{algpseudocode}%
\usepackage{listings}%
\usepackage{placeins}%

\DeclareMathOperator{\var}{var}

\newcommand{\LogS}{\mathrm{LogS}}
\newcommand{\DSS}{\mathrm{DSS}}
\newcommand{\HS}{\mathrm{HS}}
\newcommand{\ESShat}{\widehat{\mathrm{ESS}}}
\newcommand{\Rhat}{\widehat{\mathrm{R}}}
\newcommand{\Rhatmax}{\widehat{R}_{\mathrm{max}}}%

\providecommand{\tabularnewline}{\\}

\newtheorem{example}{Example}%

\newenvironment{mylist}[1]
	{\begin{list}{}
		{\settowidth{\labelwidth}{#1}
		 \setlength{\leftmargin}{\labelwidth}
		 \addtolength{\leftmargin}{\labelsep}
		 }}
	{\end{list}}

\algrenewcommand\algorithmicrequire{\textbf{Input:}}
\algrenewcommand\algorithmicensure{\textbf{Output:}}
\algnewcommand{\Initialize}[1]{%
  \State \textbf{Initialize:}
  \Statex \hspace*{\algorithmicindent}\parbox[t]{.8\linewidth}{\raggedright #1}
}

\begin{document}
\title[Parallel Bayesian cross-validation]{Bayesian cross-validation by parallel Markov chain Monte Carlo}

\author*[1]{\fnm{Alex} \sur{Cooper}}\email{alexander.cooper@monash.edu}
\author[2]{\fnm{Aki} \sur{Vehtari}}\email{Aki.Vehtari@aalto.fi}
\author[1]{\fnm{Catherine} \sur{Forbes}}\email{catherine.forbes@monash.edu}
\author[3]{\fnm{Dan} \sur{Simpson}}\email{dan@normalcomputing.ai}
\author[1,4]{\fnm{Lauren} \sur{Kennedy}}\email{lauren.a.kennedy@adelaide.edu.au}
\affil*[1]{\orgdiv{Department of Econometrics and Business Statistics}, \orgname{Monash University}, \orgaddress{ \country{Australia}}}
\affil[2]{\orgdiv{Department of Computer Science}, \orgname{Aalto University}, \orgaddress{\country{Finland}}}
\affil[3]{\orgname{Normal Computing}, \orgaddress{New York}}
\affil[4]{\orgdiv{School of Computer and Mathematical Sciences}, \orgname{University of Adelaide}, \orgaddress{\country{Australia}}}


\abstract{
Brute force cross-validation (CV) is a method for predictive assessment
and model selection that is general and applicable to a wide range
of Bayesian models. Naive or `brute force' CV approaches are often too computationally
costly for interactive modeling workflows, especially
when inference relies on Markov chain Monte Carlo (MCMC). We propose
overcoming this limitation using massively parallel MCMC.
Using accelerator hardware such as graphics processor
units (GPUs), our approach can be about as fast (in wall clock time)
as a single full-data model fit.

Parallel CV is flexible because it can easily exploit a wide range 
data partitioning schemes, such as those designed for non-exchangeable data.
It can also accommodate a range of scoring rules.

We propose MCMC diagnostics, including
a summary of MCMC mixing based on the popular potential scale reduction
factor ($\Rhat$) and MCMC effective sample size ($\ESShat$)
measures. We also describe a method for determining whether
an $\Rhat$ diagnostic indicates approximate stationarity of
the chains, that may be of more general interest for applications
beyond parallel CV. Finally, we show that parallel CV and its
diagnostics can be implemented with online algorithms,
allowing parallel CV to scale up to very large blocking designs
on memory-constrained computing accelerators.
}

\maketitle

\section{Overview}

Bayesian cross-validation \citep[CV;][]{Geisser,Vehtari2002} is a
method for assessing models' predictive ability, and is a popular
basis for model selection. Naive or `brute force' approaches to
CV, which repeatedly fits models to data subsets, are computationally demanding.
Brute force CV is especially costly when the number of folds is large and inference
is performed by Markov chain Monte Carlo (MCMC) sampling. Furthermore,
since MCMC inference must be closely supervised to identify issues
and to monitor convergence, assessing many models fits can also be labor-intensive. Consequently,
brute force CV is often impractical under conventional inference workflows
\citep[e.g.,][]{Gelman2020}.

Fast alternatives to brute force CV exist for special cases. Importance
sampling and Pareto-smoothed importance sampling \citep{Gelfand1992a,Vehtari2017}
require only a single MCMC model fit to approximate leave-one-out
(LOO) CV. However, importance sampling is known to fail when the resampling
weights have thick-tailed distributions, which is especially likely
for CV schemes designed for non-exchangeable data. Examples include 
$hv$-block CV for time series applications \citep{RACINE200039}
and leave-one-group-out (LOGO) CV for grouped hierarchical models,
where several observations are left out at the same time.
In these cases, the analyst must fall back on brute force methods.

In this paper, we show that general brute force CV by MCMC is feasible
on computing accelerator hardware, specifically on graphics processor units (GPUs).
Our method, which we call
parallel CV (PCV), includes an inference workflow and associated MCMC
diagnostic methods. PCV is not a replacement for standard inference
workflows, but rather an extension that applies after criticism
of candidate models. PCV runs inference for all
folds in parallel, potentially requiring thousands of independent
MCMC chains, and assesses convergence across all chains simultaneously
using diagnostic statistics that target the overall CV objective.
Our experiments show that PCV can estimate moderately large CV problems
on an `interactive' timescale---that is, a similar elapsed wall clock time
as the original full-data model fit by MCMC.

PCV is an application of massively parallel
MCMC, which takes advantage of recent developments in hardware and software \citep[see e.g.][]{Lao2020} and CV's embarrassingly parallel nature. Unlike the
`short chain' parallel MCMC approach which targets a single posterior, we target multiple independent posteriors concurrently
on a single computer. Other approaches include
independent chains run on separate CPU cores
and/or within-chain parallelism, both of which are
available in the Stan language \citep{TEAM2015}. Local balancing approaches 
parallel inference by generating ‘clouds’ of proposals for each MCMC
step \citep{glatt2022parallel, Gagnon2023, holbrook2023quantum}.
\cite{neiswanger2016embarrassingly} handle large datasets by targeting
a single composite posterior with distributed MCMC samplers applied to
different data subsets \citep[see also][]{vehtari2020EP,scott2016}.

In addition to parallelism, PCV further reduces computational effort
compared with naive brute force CV.
First, PCV simplifies warm-up runs by reusing information generated during full-data inference, in effect exploiting
the similarity between the full-data and partial-data CV posteriors.
Second, running chains in parallel allows early termination as soon as the required accuracy is achieved. Since Monte Carlo (MC) uncertainty is usually small relative to the irreducible CV epistemic uncertainty (see section \ref{subsec:cv}), applications of CV for model selection typically require only relatively short MCMC runs.
The third way is technical, and applies where online algorithms are used. These have small, stable working sets and make effective use of memory caches on modern computer architectures \citep[see e.g.][]{sciFastCode,stallings2015computer},
although we do not analyze this phenomenon in this paper.

\begin{figure*}
\centering{}\includegraphics[width=1\textwidth]{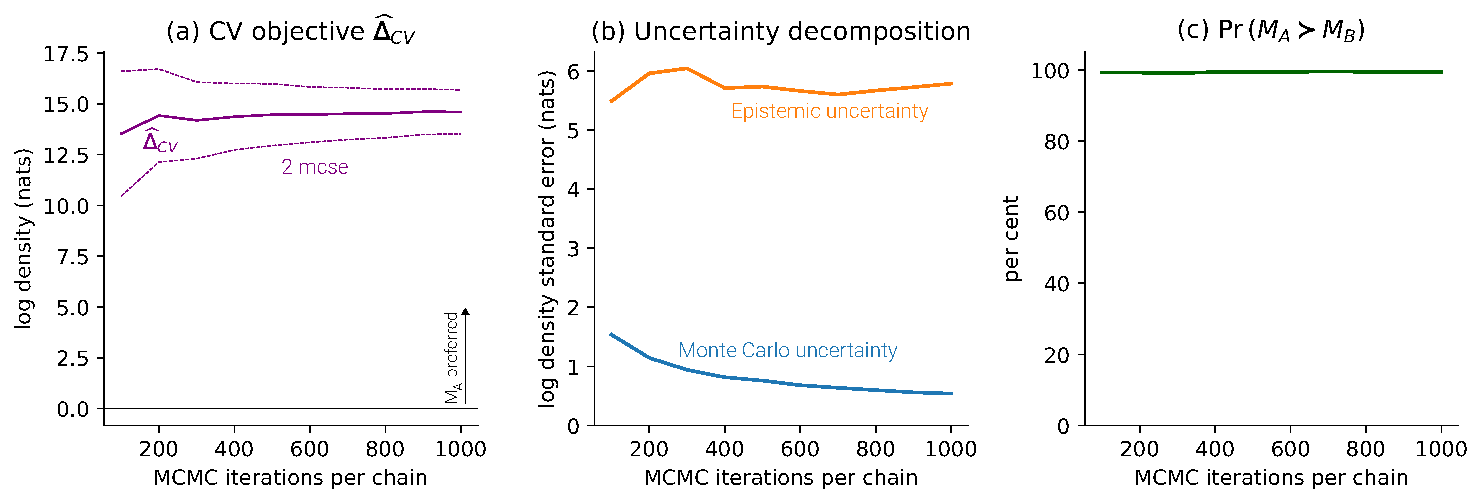}\caption{Model selection for a leave-one-group-out CV (LOGO-CV) for a toy hierarchical
Gaussian regression problem, using the log score (Example~\ref{exa:hreg}).
Model $M_{A}$ is correctly specified, including all 4 regressors,
while $M_{B}$ is misspecified and includes only 3. 
MCMC inference uses 4 chains per fold, for a total of 200 chains. Panel (a) shows
the score difference $\widehat{\Delta}$ and 2 Monte Carlo standard error (MCSE) interval as a function
of MCMC iterations per chain. Positive (negative) score values favor $M_A$ ($M_B$).
Panel (b) shows that epistemic uncertainty dominates MC uncertainty, demonstrating
the limited utility of further reducing MC uncertainty. Panel (c)
shows that $\Pr\left(M_{A}\succ M_{B}\,|\,y\right)$ quickly stabilizes close to 100\%.
See Section~\ref{subsec:pred-asst} for full details.}
\label{fig:toy-reg-sel}
\end{figure*}

Figure~\ref{fig:toy-reg-sel} previews a PCV model selection application
for a hierarchical Gaussian regression model, described in Section~\ref{subsec:pred-asst}.
The goal of this exercise is
to estimate the probability that one model predicts better than another
under the logarithmic scoring rule and a LOGO-CV design. The results
clearly stabilize after just a few hundred MCMC iterations. This is explained by the
fact that the MC uncertainty is small relative to epistemic uncertainty,
so that running the MCMC algorithm for longer confers little additional
insight.

In summary, this paper makes the following contributions to a methodological toolkit
for parallel CV:
\begin{itemize}
\item A practical workflow for fast, general brute force CV on
computing accelerators (Section~\ref{sec:pbf-CV});
\item A massively parallel sampler and associated diagnostics for PCV,
including online estimators for all estimators and diagnostics
discussed in this paper (Section~\ref{sec:impl}); and
\item A measure of MC and epistemic uncertainty \citep{Sivula2020a}, the effective sample
size ($\ESShat$; \cite{Robert2004}), and a measure of mixing based
on the $\Rhat$ statistic \citep{Gelman1992,Vehtari2020} (Section~\ref{sec:diagnostics}).
\item Examples with accompanying software implementations (Section~\ref{sec:examples} and supplement).
\end{itemize}

\section{Background}

This section provides a brief overview of predictive model assessment
and MCMC-based Bayesian inference.
Consider an observed data vector $y\sim p_{\mathrm{true}}$, where
$p_{\mathrm{true}}$ denotes some `correct' but unknown joint data distribution.
Suppose an analyst fits some Bayesian model $M$ for the purpose of
predicting unseen realizations $\tilde{y}$ from $p_{\mathrm{true}}$. Having fit
the posterior distribution $p\left(\theta\,|\,y,M\right)$, the predictive
density $f_{M}$ is a posterior-weighted mixture, 
\begin{equation}
f_{M,y}\left(\tilde{y}\right)=\int p_{y}\left(\tilde{y}\,|\,\theta,M\right)p\left(\theta\,|\,y,M\right)d\theta.\label{eq:pred-dens}
\end{equation}
With the predictive $f_{M,y}$ in hand, two natural question arise.
First, how well does $f_{M,y}$ predict unseen observations (out of
sample)? Second, if multiple models are available, which one predicts
better?

\subsection{Predictive assessment\label{subsec:pred-asst}}

If the true data distribution $p_{\mathrm{true}}\left(\tilde{y}\right)$
were somehow known, one could assess the predictive performance of $f_{M,y}$ directly
using a scoring rule. A scoring rule $S\left(q,y\right)$ is a functional
that maps a predictive density $q$ and a realization $y$ to a numerical
assessment of that prediction. A natural summary of the performance
of $f_{M,y}$ is the expected score,
\begin{equation}
S_{M,y}=\int p_{\mathrm{true}}\left(\tilde{y}\right)S\left(f_{M,y},\tilde{y}\right)\,\mathrm{d}\tilde{y}.\label{eq:pred-perf-rep}
\end{equation}
Furthermore, it is straightforward to use $S_{M,y}$ as a basis for
model selection. For a pairwise model comparison between candidate
models, say $M_{A}$ and $M_{B}$, a simple decision rule relies only
on the sign of the difference 
\begin{equation}
\Delta=S_{M_{A},y}-S_{M_{B},y}.\label{eq:model-sel}
\end{equation}
Positive values of $\Delta$ indicate $M_{A}$ is preferred to $M_{B}$
under $S$, (an event we denote $M_{A}\succ M_{B}$) and
vice-versa.

Ideally, the choice of scoring rule would be tailored to the application
at hand. However,
in the absence of an application-driven loss function, generic scoring
rules are available. By far the most commonly-used scoring rule is
the log predictive density $\LogS\left(f,y\right):=\log f(y)$, which
has the desirable mathematical properties of being local and strictly
proper \citep{Gneiting2007}. $\mathrm{LogS}$ also has deep connections
to the statistical concepts of KL divergence and entropy \citep[see, e.g.,][]{Dawid2014a}.
While we focus on $\LogS$, it is worth noting that $\LogS$ has drawbacks
too: it requires stronger theoretical conditions to reliably estimate
scores using sampling methods and can encounter problems with models
fit using improper priors \citep{Dawid2015a}. More stable results
can be obtained by alternative scoring rules, albeit at the cost of
statistical power \citep{Kruger2020}. To demonstrate the flexibility
of PCV in Appendix~A we briefly discuss the
use of alternative scoring rules, the Dawid-Sebastini score ($\DSS$;
\citealp{Dawid1999a}) and Hyvärinen score ($\HS$; \citealp{hyvarinen2005estimation}).

Throughout the paper, we illustrate ideas using the following example,
with results in Figure~\ref{fig:toy-reg-sel}.
\begin{example}[Model selection for a grouped Gaussian regression]
\label{exa:hreg}Consider the following regression model of grouped
data:
\begin{equation}
y_{ij}\,|\,\alpha_{j},\beta,\sigma_{y}^{2}\overset{ind}{\sim}\mathcal{N}(\alpha_{j}+x_{j[i]}^{\top}\beta,\sigma_{y}^{2}),
\end{equation}
for $i=1,\dots,N_{j},\quad j=1,\dots,J.$ 
The group effect prior is hierarchical, 
\begin{equation}
\alpha_{j}|\mu_{\alpha},\sigma_{\alpha}^{2}\overset{iid}{\sim}\mathcal{N}(\mu_{\alpha},\sigma_{\alpha}^{2}),\quad j=1,\dots,J,
\end{equation}
where $\mu_{\alpha}\sim \mathcal{N}\!(0,1)$ and
$\sigma_{\alpha}\sim \mathcal{N}^{+}\!(0,10)$, the half-normal
distribution with variance 10 and positive support.
The remaining priors are $\sigma_{y}\sim \mathcal{N}^{+}\!(0,10)$ and
$\beta\sim \mathcal{N}\!(0,I)$.
Consider two candidate models distinguished by their group-wise explanatory variables
$x_{j}$. Model $M_{A}$ is correctly, while model $M_{B}$ is missing
one covariate. To assess an observation $\tilde{y}$ from group $j'$ 
with explanatory variables $\tilde{x}_{j'}$ with respect to $\LogS$,
the predictive density is
\begin{align}
f_{M,y}\left(\tilde{y}\right)&=\int\mathcal{N}\!\left(\tilde{y}\,|\,\alpha_{j'}+\tilde{x}_{j'}^{\top}\beta,\sigma_{y}^{2}\right)p\left(\alpha_{j'}\,|\,y, M\right)\notag\\
&\qquad p\left(\beta,\sigma_{y}^{2}\,|\,y,M\right) d\alpha_{j'}\,d\beta\,d\sigma_{y}^{2}.\label{eq:hier-reg-pred}
\end{align}
Here and for the remainder of the paper, $\mathcal{N}\!\left(x\,|\,\mu,\sigma^2\right)$ denotes the density of
$\mathcal{N}\!\left(\mu,\sigma^2\right)$ evaluated at $x$. In the case where group $j'$ does not appear in the training data
vector $y$, the marginal posterior density $p\left(\alpha_{j'}\,|\,y,M\right)$
that appears in \eqref{eq:hier-reg-pred} is defined using the posterior
distributions for $\mu_{\alpha}$ and $\sigma_{\alpha}$, as
\begin{equation}
\int p\left(\alpha_{j'}|\mu_{\alpha},\sigma_{\alpha}\right)p\left(\mu_{\alpha},\sigma_{\alpha}\,|\,y,M\right)d\mu_{\alpha}d\sigma_{\alpha}.\label{eq:group-marg}
\end{equation}
We simulate $J=50$ groups with $N_j=5$ observations
per group. The elements of the $J\times 4$ matrix $X$ 
is simulated using $\mathcal{N}(0,10)$ variates. The true values of
$\alpha$, $\beta$ and $\sigma^2$ used in the simulation are drawn from
the priors. We use $L=4$ chains and $K=J=50$ folds, a total of 200 chains.
\end{example}

\subsection{Cross-validation\label{subsec:cv}}

In practice the true process 
$p_{\mathrm{true}}\left(\tilde{y}\right)$ is not known, so \eqref{eq:pred-perf-rep}
cannot be computed directly.
Rather, CV approximates \eqref{eq:pred-perf-rep} using only the observed
data $y$ instead of future data $\tilde{y}$. CV proceeds by repeatedly
fitting models to data subsets, assessing the resulting model predictions
on left-out data. A CV scheme includes the choice of scoring rule and
a data partitioning scheme that divides into pairs of mutually disjoint
test and training sets $\left(\mathsf{test}_{k},\mathsf{train}_{k}\right)$,
for $k=1,\dots,K$. The resulting partial-data posteriors $p\left(\cdot\,|\,y_{\mathsf{train}_{k}},M\right)$
can be viewed as $K$ random perturbations of the full-data posterior
$p\left(\cdot\,|\,y,M\right)$.

Popular CV schemes include leave-one-out (LOO-CV) which drops a single
observation each fold, and $K$-fold which divides the data into $K$
disjoint subsets. However, for non-exchangeable data, CV schemes often
need to be tailored to the underlying structure of the data, or to
the question at hand. For example, time-series, spatial, and spatio-temporal
applications can benefit from specifically tailored partitioning schemes
\citep[see e.g.,][]{roberts2017cross,Cooper2023,Mahoney2023}. For
some data structures the resulting $K$ can be particularly large,
such as for LOO-CV and $h(v)$-block CV.

There may be several appropriate CV schemes for a given candidate
model and dataset. The CV scheme should also reflect the nature
of the generalization required. For example, in hierarchical models
with group effects (Example~\ref{exa:hreg}), LOGO-CV measures the 
model's ability to generalize to unseen
data groups, while non-groupwise schemes applied to the same model
and data would characterize predictive ability for the current set
of groups only.

We focus on two  CV objectives. First, the CV score $\widehat{S}_{M}$
is an estimate of predictive ability $S_{M}$. It is constructed as
a sum over all $K$ folds,
\begin{equation}
\widehat{S}_{M}=\sum_{k=1}^{K}S\left(f_{M,k},y_{\mathsf{test}_{k}}\right),\label{eq:cv-score}
\end{equation}
where $f_{M,k}$ denotes the model $M$ predictive constructed using the
posterior $p\left(\theta\,|\,y_{\mathsf{train}_k}\right)$. The quantity
$\widehat{S}_{M}$ can be viewed as a Monte Carlo estimate of $S_{M}$,
up to a scaling factor. Second, a CV estimate for the model selection
objective $\Delta$ in \eqref{eq:model-sel} is the sum of the $K$
differences,
\begin{align}
\widehat{\Delta} & :=\widehat{S}_{M_{A}}-\widehat{S}_{M_{B}}\notag\\
&=\sum_{k=1}^{K}\left[S\left(f_{M_{A},k},y_{\mathsf{test}_{k}}\right)-S\left(f_{M_{B},k},y_{\mathsf{test}_{k}}\right)\right]\notag\\
&=:\sum_{k=1}^{K}\widehat{\Delta}_{k}.\label{eq:sel-cv}
\end{align}

Herein we use the generic notation $\widehat{\eta}=\sum_{k=1}^{K}\widehat{\eta}_{k}$
to denote the CV objective, whether it be $\widehat{S}_{M}$ or $\widehat{\Delta}$.

\subsection{Epistemic uncertainty}\label{subsec:epistemic}

The MC estimators in \eqref{eq:cv-score} and \eqref{eq:sel-cv} are random quantities
that are subject to sampling variability.
The associated predictive assessments of out-of-sample model performance
are therefore subject to uncertainty \citep{Sivula2020a}.
Naturally, there would be no uncertainty at all if $p_{\mathrm{true}}$ were fully known or if an infinitely large dataset were available. But since \eqref{eq:cv-score} and \eqref{eq:sel-cv}
are estimated from finite data, an assessment of the 
uncertainty of these predictions is useful for interpreting CV results \citep[][\S 2.2]{Sivula2020a}. We call this random error \emph{epistemic uncertainty}.

It can be helpful to view as-yet-unseen data as missing data. 
CV imputes that missing data with finite number of left-out observations. The
epistemic uncertainty about the unseen future data can be regarded as
uncertainty from the imputation process.

For a given dataset, epistemic uncertainty is irreducible in the sense that it cannot 
be driven to zero by additional computational effort
or the use of more accurate inference methods. This contrasts with Monte Carlo uncertainty
in MCMC inference, discussed in the next subsection. The limiting factor is the
information in the available dataset.

We adopt the following popular approach to modeling epistemic uncertainty.
First, we regard the individual contributions 
$\left(\widehat\eta_k\right)$ to $\widehat\eta$ as exchangeable, drawn from a large population 
with finite variance \citep{Vehtari2002,Sivula2020a}. That is, we presume that $K$ is reasonably large.
Then, $\widehat\eta$ will satisfy a central limit theorem (CLT)
so that a normal approximation for the out-of-sample predictive performance of
$\widehat\eta$ is appropriate.

In particular, for model selection applications we are interested in the probability that $M_A$
predicts better than $M_B$, which we denote 
$\Pr\left(M_{A}\succ M_{B}\,|\,y\right)$. We approximate
\begin{equation}
\Pr\left(M_{A}\succ M_{B}\,|\,y\right)\approx\Phi\left(\frac{\widehat{\Delta}}{\sqrt{K\widehat{\sigma}_{\widehat{\Delta}}^{2}}}\right),\label{eq:norm-approx}
\end{equation}
where $\widehat{\sigma}_{\widehat{\Delta}}^{2}=\frac{1}{K-1}\sum_{k=1}^{K}\left(\widehat{\Delta}_{k}-\widehat{\Delta}/K\right)^{2}$
is the sample variance of the contributions to \eqref{eq:sel-cv} and
$\Phi$ denotes the standard normal cdf.

There are other ways to model
$\Pr\left(M_{A}\succ M_{B}\,|\,y\right)$, such as the Bayesian Bootstrap
\citep{Rubin,Sivula2020a}. We prefer the normal approximation
in this setting because it performs well \citep{Sivula2020a} and is simple to approximate with 
online estimators, making it particularly useful on parallel hardware.

\subsection{MCMC inference\label{subsec:MCMC-inference}}

MCMC is by far the dominant method for conducting Bayesian inference.
It characterizes posterior distributions with samples from a stationary
Markov chain, for which the invariant distribution is the target posterior.
MCMC sampling algorithms are initialized with a starting parameter
value $\theta^{\left(0\right)}$, usually in a region of relatively
high posterior density. Then, an MCMC algorithm is used
to sequentially draw a sample $\theta^{\left(1\right)},\theta^{\left(2\right)},\dots$ from the chain.
This sample can be used to construct Monte Carlo estimators for functionals
$g\!\left(\theta\right)$ such as the predictive density $f_{M,y}\!\left(\tilde y\right)$
that appears in \eqref{eq:pred-dens}. Many MCMC algorithms are available:
\citet{Gelman2014} provide a general overview.

For brute-force CV applications, posteriors are separately estimated
for each model and fold. Furthermore, typical inference workflows
for MCMC inference call for draws from $L>1$ independent chains
targeting the same posterior. For model $M$, fold $k$, and chain
$\ell$, denote the sequence of $N$ MCMC parameter draws $\theta_{M,k,\ell}^{\left(0\right)},\theta_{M,k,\ell}^{\left(1\right)},\cdots,\theta_{M,k,\ell}^{\left(N\right)}$,
so that the expectation $\mathbb{E}\left[g\left(\theta\right)\,|\,M,y_{\mathsf{train}_k}\right]$
can be estimated for each model $M$ and fold $k$ as
\begin{equation}
\widehat{g}_{M,k}^{\left(N\right)}=\frac{1}{LN}\sum_{\ell=1}^{L}\sum_{n=1}^{N}g\left(\theta_{M,k,\ell}^{\left(n\right)}\right).\label{eq:est-e}
\end{equation}

The main assumption needed is quite standard (see e.g. \citealp{Jones2006}): a CLT
for $\widehat{g}_{M,k}^{\left(N\right)}$, so that
\begin{align}
&\sqrt{N}\left(\widehat{g}_{M,k}^{\left(N\right)}-\mathbb{E}\left[g\left(\theta\right)\,|\,M,y_{\mathsf{train}_k}\right]\right)\notag\\
&\qquad\qquad\overset{d}{\longrightarrow}\mathcal{N}\left(0,\frac{\sigma_{\widehat{g}_{M,k}}^{2}}{L}\right)\ \text{as}\  N\to\infty.\label{eq:clt}
\end{align}

Because MCMC parameter draws are auto-correlated, a naive estimate
of the uncertainty associated with \eqref{eq:est-e} from these draws
will be biased. Instead, a Monte Carlo standard error (MCSE) estimator
should be used. We use MCSE estimators based on batch means \citep{Jones2006}
because these can be efficiently implemented on accelerator
hardware (see Appendix~A). Let the chain length
be a whole number representing $a$ batches each of $b$ samples, so that
$N=ab$. Then the $h$th batch mean is given by
\begin{equation}
\bar{g}_{M,k,\ell}^{\left(h\right)}=\frac{1}{b}\sum_{n=\left(h-1\right)b+1}^{hb}g\left(\theta_{M,k,\ell}^{\left(n\right)}\right).\label{eq:bmean}
\end{equation}
The MCSE $\sigma_{\hat{g}_{M,k}}/\sqrt{LN}$ can then be computed using
the sample variance of the $\bar{g}_{M,k,\ell}^{\left(h\right)}$s
across all $L$ chains for model $M$ and fold $k$, where
\begin{equation}
\hat{\sigma}_{\widehat{g}_{M,k}}^{2}=\frac{b}{La-1}\sum_{\ell=1}^{L}\sum_{h=1}^{a}\left(\bar{g}_{M,k,\ell}^{\left(h\right)}-\widehat{g}_{M,k}^{\left(N\right)}\right)^{2}.\label{eq:bmvar}
\end{equation}
There is a large literature on estimators for $\sigma_{\widehat{g}_{M,k}}^{2}$,
but we use  \eqref{eq:bmvar} since it is simple and it performed well
in our experiments. The batch size $b$ is a hyper-parameter to
be chosen before inference starts, and large
enough for the $Y_{k}$ to be approximately independent. Where the
MCMC chain length is known (in the case where a data-dependent stopping rule
is not used), asymptotic arguments suggest $b=\left\lfloor \sqrt{NL}\right\rfloor $,
for which a rough guess can be made \emph{a priori} (see for instance
\citealp{Jones2006} for a discussion).

To reliably fit Bayesian models, the inference workflow needs to include
careful verification of model fit. MCMC algorithms must also be carefully
checked for pathological behaviors and monitored for convergence so
that inference can be terminated \citep{Gelman2020}. Most workflows
are oriented toward parameter inference, ensuring that the samples
adequately characterize the desired posterior distribution $p\left(\theta\,|\,y\right)$.
Assessing convergence effectively amounts to verifying that (a) each
posterior's chains are correctly mixing, and (b) the sample size is
large enough to characterize the posterior distribution to the desired
degree of accuracy. To support these assessments, several diagnostic
statistics are available \citep{Roy2020}. We describe two diagnostic
statistics specifically adapted for parallel MCMC in
Section~\ref{sec:diagnostics}.

\section{Parallel cross-validation\label{sec:pbf-CV}}

In this section we describe the proposed PCV workflow. 
PCV aims to make brute force CV feasible by reducing computation (wall clock)
time as well as the analyst's time spent checking diagnostics. Of
course, given unlimited computing power and effort, an analyst could
simply implement Bayesian CV by sequentially fitting each CV fold
using MCMC, checking convergence statistics for each, then constructing the
objective $\widehat{\eta}$ using \eqref{eq:cv-score} or \eqref{eq:sel-cv}.
This is computationally and practically infeasible for CV designs
with large $K$.

Many existing applications of massively parallel MCMC target just a single
posterior. In contrast, PCV simultaneously estimates
multiple posteriors using parallel samplers that execute in lock-step. Under our approach,
the algorithm draws vectors of combined parameter vectors representing
all chains for all folds. Estimating $\hat{\eta}$ by brute force
MCMC requires samples from $\mathcal{O}\left(KL\right)$ parallel
independent MCMC chains. In some cases, posteriors for different models
can be sampled in parallel, too (see Appendix~B).

The main challenge that must be solved for estimating all $K$ posteriors
(or $2K$ posteriors for a pairwise model comparison) is that model
criticism, diagnostics, and convergence checking must be applied to
all posteriors under test. However, since full model criticism and
checking should be performed on the full-data models anyway, for which
the partial-data posteriors are simply random perturbations, we argue
that model criticism need only be applied to the full-data candidate
models. Furthermore, we can sharply reduce the computational effort
required for sampling by using information from full-data inference
to initialize the partial-data inference (Section~\ref{subsec:warmup}).

We propose the following extension to conventional Bayesian inference workflows:
\begin{mylist}{00.00.0000}
\item [{\textbf{Step~1.}}] \textbf{Full-data model criticism and inference}. Perform
model criticism on candidate model(s) using the full data set \citep{Gelman2020},
revising candidate models as necessary, and obtain
MCMC posterior draws for each model;
\item [{\textbf{Step~2.}}] \textbf{Parallel MCMC warmup} (see Section~\ref{subsec:warmup}). Initialize $L$
parallel chains for each of $K$ folds using random draws from the full-data
MCMC draws obtained in Step 1. Run short warm-up chains in parallel
and discard the output;
\item [{\textbf{Step~3.}}] \textbf{Parallel sampling}.\textbf{ }Run the $\mathcal{O}(KL)$ parallel MCMC chains for $N$
iterations, accumulating statistics 
required to evaluate $\widehat{\eta}$ and associated uncertainty measures (see Section~\ref{sec:uncert});
and
\item [{\textbf{Step~4.}}] \textbf{Check convergence}. Compute and check parallel
inference diagnostics (Section~\ref{sec:diagnostics}),
and if necessary adjust inference settings and repeat.
\end{mylist}

\subsection{Efficient MCMC warmup\label{subsec:warmup}}

General-purpose MCMC sampling procedures typically begin with a warmup phase. The warmup serves two goals: (i) it reduces MCMC estimator bias due to initialization, and (ii) adapts tuning parameters of the sampler. Warm-up procedures aim to ensure the distribution of the initial chain values $\theta_\ell^{(0)}$ is close to that of the target posterior.
An example is Stan's window adaptation algorithm \citep{TEAM2015},
designed for samplers from the Hamiltonian Monte Carlo \citep[HMC;][]{Neal2011} family. Hyperparameter choices are algorithm-specific: for example, 
HMC requires a step size, trajectory length,
and inverse mass matrix.

The warm-up phase is computationally costly. It would be especially costly and
time consuming to run complete, independent tuning procedures for
each fold in parallel to obtain kernel hyperparameters and initial conditions
$\mathcal{O}\left(KL\right)$  for each fold. In addition, running
many independent tuning procedures can be unreliable. Tuning procedures are stochastic
in nature, and as the number of chains
increases, the probability of initializing at least one chain with problematic starting
conditions increases. An example of such a starting condition is a parameter draw
far in a region of the parameter space that leads to numerical problems and a `stuck chain'.

Instead of running independent warm-up procedures for each fold, 
we propose re-using the warm-up results from the full-data model for each CV fold,
under the assumption that the full-data and partial-data posteriors are close. 
This assumption seems reasonable if the folds are similar enough to the 
full-data model for CV to be interpretable as a predictive assessment of the full-data 
model (Section~\ref{subsec:cv}).

Under this approach, each fold's MCMC kernel uses the same inference tuning parameters (e.g. step size and trajectory length) as the full-data model. Starting positions are randomly drawn from the full-data
posterior MCMC sample. To ensure distribution of the
starting conditions are close to
the fold model's posterior distribution, 
PCV then simulates and discards a very
short warm-up sample.

\subsection{Estimating uncertainty}\label{sec:uncert}

Practical CV applications require estimates of the uncertainty of
$\widehat{\eta}$ estimates. 
Both MC and epistemic uncertainty contribute
to the variability in $\widehat{\eta}$. 

We estimate epistemic uncertainty by applying the normal approximation described in Section~\ref{subsec:epistemic}.
For model selection application, we substitute fold-level estimates $\left(\widehat\Delta_k\right)$ into \eqref{eq:norm-approx}.


MC uncertainty can be estimated using an extension of standard methods described in Section~\ref{subsec:MCMC-inference}. Because each fold is estimated using independent
MCMC runs, the overall MC uncertainty for $\widehat{\eta}$ is simply
the sum of the MC variance of each fold's contribution. When an estimated scoring
rule may be represented as a smooth function of an ergodic mean
(like $\LogS$), $\sigma_{\hat{\eta}}^{2}$ can be estimated using the delta method.
In the case of $\LogS$, we have
\begin{equation}
\sigma_{\hat{\eta}}^{2}=\sum_{M\in\mathcal{M}}\sum_{k=1}^{K}\sigma_{S_{M,k}}^{2}\approx\sum_{M\in\mathcal{M}}\sum_{k=1}^{K}\frac{\sigma_{\hat{f}_{M,k}}^{2}}{\left(\hat{f}_{M,k}\right)^{2}},\label{eq:mc-var}
\end{equation}
where summation over models reflects the fact that the MC error for a 
difference is the sum of the error for both terms. The MCSE for 
$\hat{\eta}$ is then $\mathrm{MCSE}_{\hat{\eta}}=\sigma_{\hat{\eta}}/\sqrt{LN}.$
Independence of the contributions to the overall MC error is helpful
for producing accurate estimates quickly.

Even when $\eta$ and its associated standard error can theoretically
be estimated using MC estimators (for instance if its first two moments
are finite), in practice theoretical conditions may not be
enough to prevent numerical problems during inference.
A common cause of numerical overflow is the presence of outliers
that fall far in the tails of predictive distributions.

In model selection applications, it is typical for MC uncertainty
to be an order of magnitude smaller than epistemic uncertainty (see e.g.
Figure~\ref{fig:toy-reg-sel}). This discrepancy implies that, provided that the chains have mixed, long
MCMC runs are not usually necessary in model selection applications, since additional effort applied to MCMC
sampling will not meaningfully improve the accuracy of $\widehat{\eta}$.
An overall measure that provides a single picture of uncertainty is also
useful because the MCMC efficiency of
individual folds can vary tremendously (see Section~\ref{subsec:ess} and 
Figure~\ref{fig:grouped-reg-ess}).

Since PCV applications typically require only relatively small $N$
to make MC uncertainty insignificant compared to epistemic uncertainty,
we do not propose a stopping rule for of the type discussed
by \citet{Jones2006}. In most cases these rules are justified by
asymptotic arguments (i.e. for large $N$), whereas in our examples
$N$ is on the order of $10^{3}$.

\subsection{Implementation on accelerator hardware\label{sec:impl}}

Massively parallel samplers are able to take advantage of modern computing
accelerators, which can offer thousands of compute units per device.
Accelerators typically offer more cost-effective throughput, measured in floating-point 
operations per second (FLOPS), than conventional CPU-based workstations. However,
despite their impressive parallel computing throughput and cost per
FLOPS, the design of computing accelerators impose heavy restrictions
on the design of inference algorithms.

Programs with heavy control flow beyond standard linear algebra operations
tend to be inefficient, including those commonly used in Bayesian
inference such as sorting large data vectors. In addition, accelerators
have limited onboard memory for storing and manipulating draws generated
by MCMC samplers. The need to transfer MCMC draws to main memory for
diagnostics and manipulation would represent a significant performance
penalty. These problems could be alleviated if all analysis steps
could be conducted on the accelerator device, within its memory limits.

The Hamiltonian Monte Carlo \citep[HMC;][]{Neal2011} sampling algorithm
lends itself well to implementation on accelerator hardware. To implement
PCV, we augment the HMC kernel with an additional parameter representing
the fold identifier. This allows the unnormalized log joint density---and
its gradient via automatic differentiation---to select the appropriate
data subsets for each chain. 
HMC is suitable for parallel operations because its integrator follows
a fixed trajectory length at each MCMC iteration. The lack of a dynamic
trajectory allows HMC to be vectorized across a large number
of parallel chains, with each evolving in lock-step. In contrast,
efficient parallel sampling is extremely difficult with dynamic algorithms
such as the popular No-U-Turn Sampler \citep[NUTS;][]{Hoffman2011a}.
Assessing chain efficiency (Section~\ref{subsec:ess}) is more important
with samplers with non-adaptive step size like HMC, since samples
tend to be more strongly auto-correlated than for adaptive methods,
delivering fewer effective draws per iteration.

A further constraint inherent to computing accelerators is the size of the
device's on-board memory. Limited accelerator memory means it is usually 
infeasible to store all MCMC parameter draws from all chains for later analysis,
which requires only $\mathcal{O}\left(KLN\dim\left(\theta\right)\right)$ memory.
This cost can be prohibitive, especially when $\dim\left(\theta\right)$ is large.
However, it is often feasible to store draws required to construct the objective,
that is the univariate draws required to estimate $\hat\eta$, reducing the memory
requirement to $\mathcal{O}\left(KLN\right)$. This approach is very simple to implement
(see Algorithm~\ref{alg:simple-sampler}).

Where accelerator device memory is so constrained that even the draws for $\hat\eta$
cannot be stored on the accelerator, online algorithms are available.
The memory footprint for such algorithms does not depend on the chain length $N$.
However, as Algorithm~A2 in the supplementary material demonstrates,
the fully-online approach is significantly more complicated.
(See also the sampler implemented in Appendix~D in the supplementary material.)
While the online approach is very memory efficient, 
it is also less flexible. Conventional workflows
recommend first running inference then computing diagnostics
from the resulting draws. However, under the online approach one must predetermine
which diagnostics will be run after inference, then accumulate enough information during
inference to compute those diagnostics without reference to the full set of
predictive draws (see Appendix~A).

\begin{algorithm*}
\scriptsize
\begin{algorithmic}
\Require Posterior draws $\left\{\theta^{(\mathrm{fd})}_1,\dots,\theta^{(\mathrm{fd})}_{N_\mathrm{fd}}\right\}$, MCMC kernels $\left\{\mathcal{T}_k\left(\theta,\cdot\right)\right\}_{k=1}^K$, log predictive densities $\left\{\log p\left(\tilde{y}_k\,|\,\theta\right)\right\}_{k=1}^K$, folds $K$, chains $L$, warm-up length $N_{\mathrm{wu}}$, chain length $N$
\Ensure $\hat{s}$.
\Initialize {$K\times L \times N$ array $\hat{s}$}
\For{$k \in 1,\dots,K$} in parallel:
\Comment{fold loop}
\For{$\ell \in 1,\dots,L$} in parallel:
\Comment{chain loop}
\State $\theta^{(k,\ell)} \gets \text{draw from }\left\{\theta^{(\mathrm{fd})}_1,\dots,\theta^{(\mathrm{fd})}_{N_\mathrm{fd}}\right\}$
\For{$i \in 1,\dots, N_{\mathrm{wu}}$} sequentially
\Comment{warm-up sampling loop}
    \State $\theta^{(k,\ell)} \gets \text{draw from } \mathcal{T}_k\left(\theta^{(k,\ell)},\cdot\right)$
    \EndFor
    \For{$i \in 1,\dots, N$} sequentially
        \Comment{sampling loop}
        \State $\theta^{(k,\ell)} \gets \text{draw from } \mathcal{T}_k\left(\theta^{(k,\ell)},\cdot\right)$
        \State $\hat{s}^{(k,\ell,i)} \gets \log p\left(\tilde{y}_k\,|\,\theta^{(k,\ell)}\right)$
        \EndFor
    \EndFor
\EndFor
\end{algorithmic}
\caption{Non-online parallel CV sampler for LogS. Superscripts index array elements.
See also Algorithm~A2 in the supplementary material.}
\label{alg:simple-sampler}
\end{algorithm*}

\section{MCMC diagnostics\label{sec:diagnostics}}

We propose that diagnostics focus on $\widehat{\eta}$ rather than fold-specific
parameters. Under our proposed workflow, the analyst will have completed
model criticism on the full-data model before attempting brute force
CV. Soundness of the full-data model strongly suggests that the CV
folds, which by construction have the same model structure and similar
data, will also behave well. However, inference of all folds should
nonetheless be monitored to ensure convergence has been reached, and
to identify common problems that may arise during computation.

A focus on the predictive quantity $\hat{\eta}$ rather than the model parameters carries several
advantages. First, it provides a single view of convergence that targets the desired output and ignores any inference problems in
irrelevant parts of the model, such as group-level random effects 
that are not required to predict the group of interest. Second,
unlike parameter convergence diagnostics, diagnostics for $\hat{\eta}$
are sensitive to numerical issues arising in the predictive components
of the model. Third, these diagnostics can be significantly cheaper
to compute than whole-parameter diagnostics, in part because 
the target is univariate or low-dimensional.

Other diagnostic statistics for massively parallel inference include
$\mathfrak n \widehat R$ \citep{Margossian2021}, which is applicable to large numbers of short
chains targeting the same posterior. In contrast, our diagnostics target a small
number of longer chains per fold, and are applied to a large number of different posteriors.

\subsection{Effective sample size}\label{subsec:ess}

An (estimated) effective sample size ($\widehat{\mathrm{ESS}}$; \citealp{Geyer1992a})
provides a scale-free measure of the information content of an autocorrelated
sample. Since MCMC samples from a single chain are not independent, an estimate of the limiting variance $\sigma^2_{\hat{g}}$ in (\ref{eq:clt}), denoted by $\widehat{\sigma}^{2}_g$, is typically greater than the usual sample variance $s_g^2$, and hence the degree of autocorrelation
must be taken into account when computing the Monte Carlo standard
error (MCSE) of estimates computed from the resulting MCMC sample. 

The standard ESS measure targets individual parameter estimates, say
$\theta_i$. Define $\ESShat_{\theta_{i}}$,
as the raw sample size adjusted by the ratio of the unadjusted sample
variance $s^{2}_{\theta_i}$ to the corresponding
MC variance $\hat{\sigma}_{\theta_{i}}^{2}$: $\ESShat_{\theta_{i}}=LNs_{\theta_{i}}^{2}/\hat\sigma_{\theta_{i}}^{2}.$
For parallel inference we use the batch means method described in Section~\ref{subsec:MCMC-inference} to estimate $\hat\sigma_{\theta_{i}}^2$,
for which online estimators are simple to implement. (For alternatives see
e.g., the review by \citealp{Roy2020}.)

\begin{figure*}
\centering{}\includegraphics[width=1\textwidth]{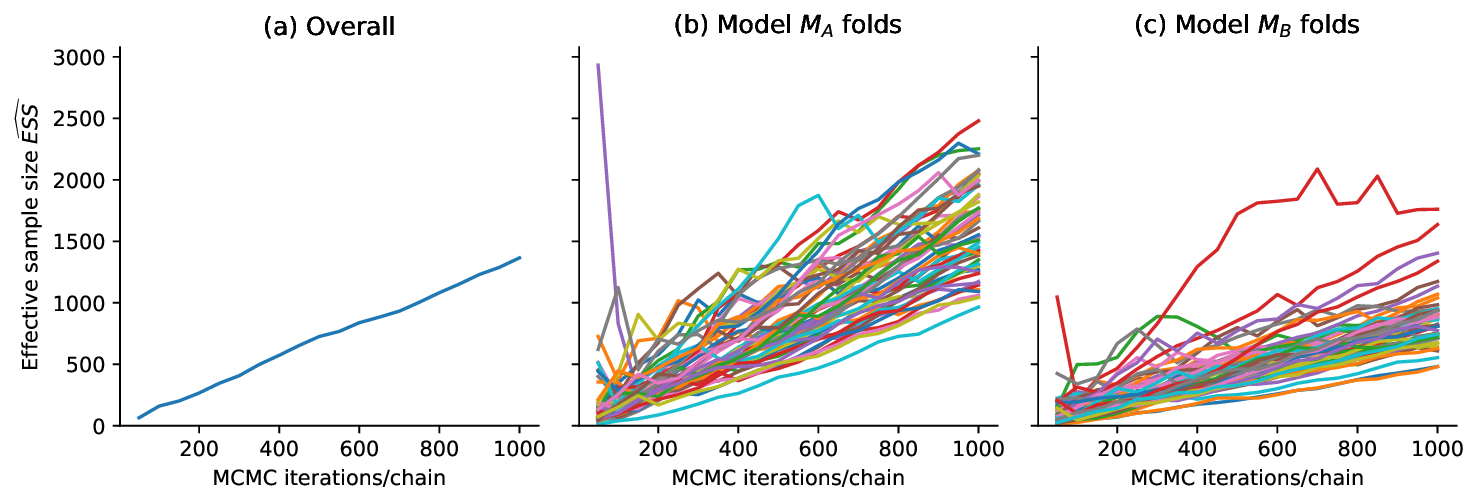}\caption{Progressive estimated effective sample size ($\ESShat$) for Example~\ref{exa:hreg} (grouped Gaussian
regression), a scale-free measure of information content for
the MCMC sample, as a function of MCMC iterations. Panel (a) shows $\ESShat$ for the overall
model selection statistic (incorporating all chains), while panels
(b) and (c) respectively show the $\ESShat$ for each fold of
models $M_{A}$ and $M_{B}$. Note the overall measure falls within
the range of all folds.}
\label{fig:grouped-reg-ess}
\end{figure*}

For PCV, an aggregate measure of the sample size 
\emph{$\ESShat$} can be
computed similarly. Define  $\ESShat=LNs_{\hat{\eta}}^{2}/\hat{\sigma}_{\hat{\eta}}^{2}$. $\ESShat$
is useful as a single scale-free measure across all folds
(see Figure~\ref{fig:grouped-reg-ess}).

\subsection{Mixing: aggregate \texorpdfstring{$\Rhatmax$}{Rhatmax}}\label{sec:rhat}

To assess mixing of the ensemble of chains, we propose a combined
measure of mixing based on the potential scale reduction
factor $\Rhat$ \citep{Gelman1992,Vehtari2020}. 
Most of the chains in the ensemble should have $\widehat{R}$s
below a suitable threshold (\citet{Vehtari2020} suggest 1.01).
In addition, we assess overall chain convergence by $\Rhatmax$,
the maximum of the $\Rhat$s measured across all folds. Below, we
describe a simple problem-specific method for interpreting $\Rhatmax$.

The canonical $\Rhat$ measure aims to assess whether the independent chains
targeting the same posterior adequately characterize the whole posterior
distribution. $\widehat R$ usually targets parameter means, but in our experiments
we found that the mean of the log score draws to be a useful $\widehat{R}$ target.
Other targets are of course possible, and wide range of other functionals appear in
the literature \citep[e.g.][]{Vehtari2020,Moins2022}.

$\Rhat$ is a scaled measure of the variance of between-chain
means $B$, a quantity that should decrease to zero as the chains
converge in distribution and become more similar. Several variants
of $\Rhat$ exist. To simplify computation on accelerators,
the simple version we use here omits chain splitting and rank-normalization (these features are described by \citealp{Vehtari2020}). 
For a given model $M$ and fold $k$, define $\Rhat$ as
\begin{equation}
\Rhat_{M,k}=\sqrt{\frac{\frac{N-1}{N}\widehat{W}_{M,k}+\frac{1}{N}\widehat{B}_{M,k}}{\widehat{W}_{M,k}}}.\label{eq:rhat}
\end{equation}
The within-chain variance $W_{M,k}$ and between-chain variance $B_{M,k}$ are, respectively
\begin{align*}
W_{M,k} & =\frac{1}{L}\sum_{\ell=1}^{L}\frac{1}{N-1}\sum_{n=1}^{N}\left(\hat{s}_{M,k,\ell}^{\left(n\right)}-\bar{s}_{M,k,\ell}\right)^{2}, \\
B_{M,k} & =\frac{N}{L-1}\sum_{\ell=1}^{L}\left(\bar{s}_{M,k,\ell}-\bar{s}_{M,k,\cdot}\right)^{2},\label{eq:WB-param}
\end{align*}
where $\hat{s}_{M,k,\ell}^{\left(n\right)}$ is the $n$th draw of $\log p(\tilde{y} | \theta)$ in chain $\ell$, $\bar{s}_{M,k,\ell}$ is the chain sample mean, and $\bar{s}_{M,k,\cdot}$
is the sample mean of parameter draws for the fold $k$ chains.

The summary mixing measure is then
\begin{equation}
\widehat{R}_{\mathrm{max}}=\max_{M\in\mathcal{M},k=1,\dots,K}\widehat{R}_{M,k}.\label{eq:rhat-max}
\end{equation}
Since all the $\Rhat_{M,k}$ tend to 1 as chains converge and $K$ is fixed, it follows
that $\Rhatmax$ tends to 1 as all posterior chains converge.

However, while $\hat{R}_{\max}$ has the same limiting value as $\Rhat$,
it is not at all clear that the broadly accepted threshold of $\Rhat<1.01$
\citep{Vehtari2020} for a single posterior is an appropriate indicator that all 
folds have fully mixed. Each $\Rhat_{M,k}$ is a stochastic quantity,
and the extremum statistic $\Rhatmax$ is likely to be large relative
to the majority of chains.

\begin{figure*}
\begin{centering}
\includegraphics[width=1\textwidth]{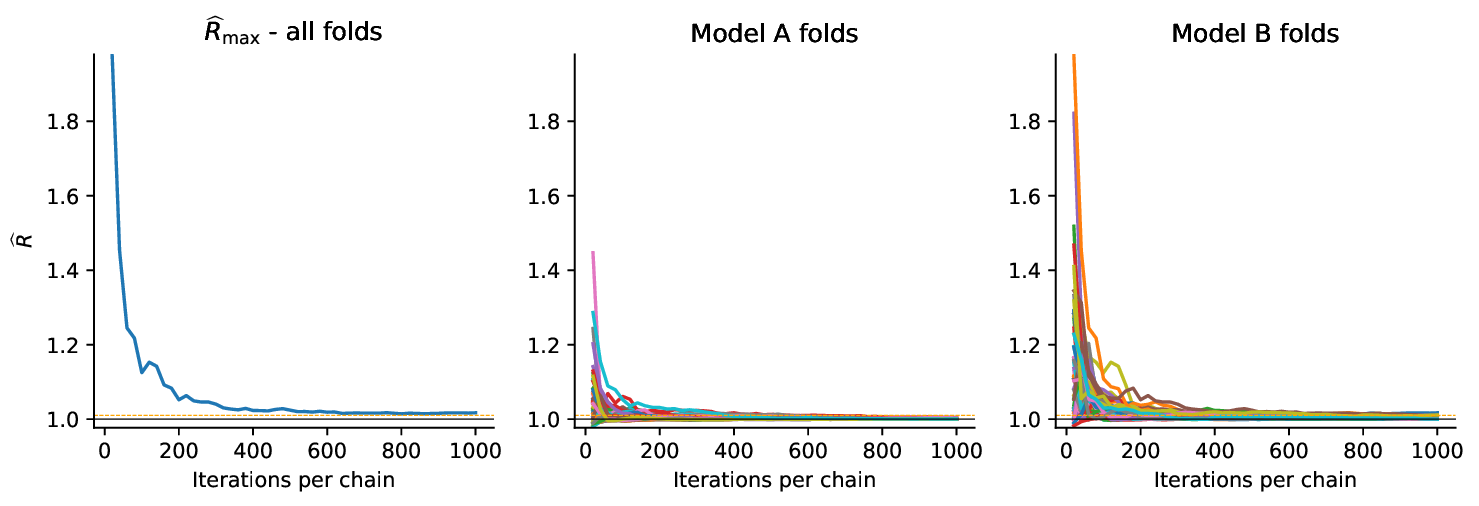}
\par\end{centering}
\caption{Progressive $\Rhat$ measures as a function of iterations/chain
for the toy grouped regression model comparison. Note that the 1.01 threshold
\citep{Vehtari2020} is exceeded by at least one chain, and hence by $\Rhatmax$,
for most of the 1,000 iterations/chain plotted in this diagram.}
\label{fig:rhat-gauss}
\end{figure*}

Figure~\ref{fig:rhat-gauss} compares $\Rhatmax$ with the
$\Rhat$ computed for each fold of both regression models $M_{A}$
and $M_{B}$. Recall from Figure~\ref{fig:toy-reg-sel} that that
the $\widehat{\eta}$ estimates stabilize after a few hundred MCMC
iterations per chain. However, well beyond this point, $\Rhatmax$ 
exceeds the conventional convergence threshold for $\Rhat$ of 1.01
\citep{Vehtari2020}.

To estimate an appropriate benchmark for $\Rhatmax$ for a given problem,
we propose the following simulation-based procedure. This procedure empirically
accounts for the autocorrelation in each chain, without the need to
model the behavior of each fold's posterior, and with only minimal additional
computation. This approach is conceptually
similar to the block bootstrap, and it directly accounts for the autocorrelation
in each fold's MCMC chains.

\begin{figure*}
\begin{centering}
\includegraphics[width=1\textwidth]{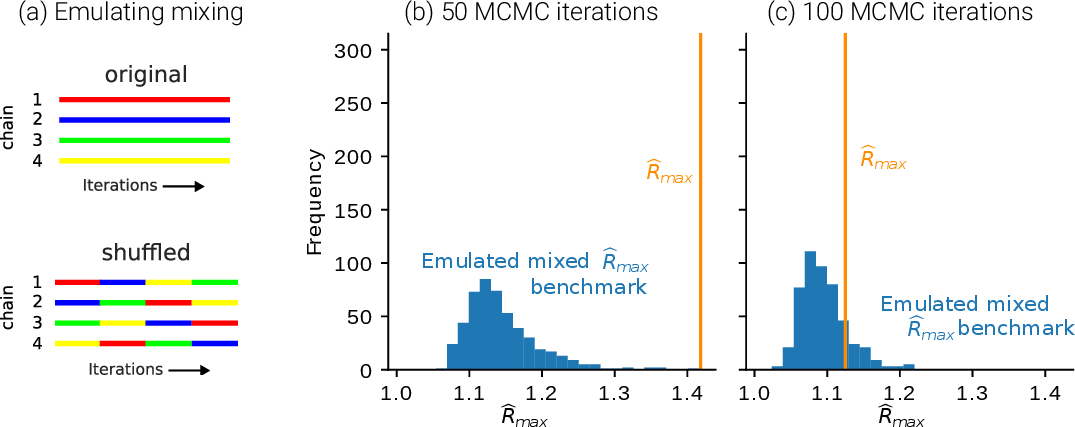}
\par\end{centering}
\caption{Two values of $\Rhatmax$ for the toy regression problem,
compared with emulated mixed $\Rhatmax$ benchmark draws, 
computed by block-shuffling
chains. Panel (a) shows a stylized shuffling scheme, where chains
are broken into contiguous blocks and recombined by shuffling with
replacement. Panels (b) and (c) show $\Rhatmax$ estimates
at 50 and 100 parallel MCMC iterations, respectively (vertical
line), alongside histograms of 500 shuffled $\Rhatmax$
draws for comparison. 5 blocks were used. In this example, we conclude
that the parallel chains have not converged after 50 iterations, but they
have after 100.}
\label{fig:shuffle}
\end{figure*}

Suppose (hypothetically) that all chains are well-mixed, so that the
mean and variance of any chain should be roughly the same.
In that case, if we also assume that autocorrelation is close to zero
within the block size, then computing $\Rhat$ should not be greatly affected
if blocks of each chains are `shuffled' as shown in Panel (a) of
Figure~\ref{fig:shuffle}. To construct an estimate of the likely
range of $\Rhat$ values under the assumption that the chains have mixed,
we simply repeatedly compute $\Rhat$ from a large sample
of shuffled draws. Rather than adopt a threshold based on an arbitrary
summary statistic of the shuffled draws as a single benchmark (say an upper quantile),
we instead simply present the draws as a histogram for visual comparison.

\begin{figure*}
\begin{centering}
\includegraphics[width=1\textwidth]{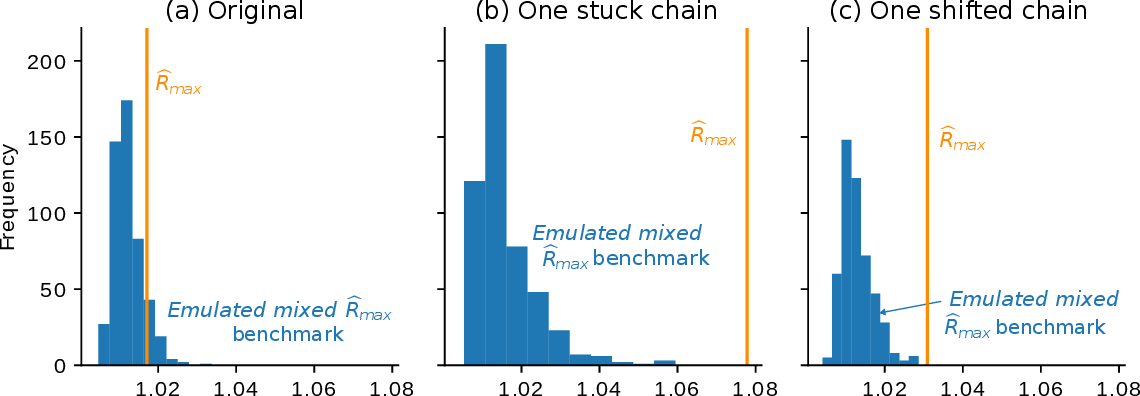}
\par\end{centering}
\caption{Artificially constructed pathological inference examples, detected by $\Rhatmax$.
Panel (a) shows the original comparison in Figure~\ref{fig:shuffle},
after 1,000 iterations. In Panel (b) a single chain for
one of the folds has been fixed to a constant value (its first draw).
In Panel (c) a single chain for one of the folds has been shifted by 5 units.
In both panels (b) and (c), $\Rhatmax$ lies to the right
of the histogram of 100 emulated stationary $\Rhatmax$ draws,
estimated using 5 blocks.}
\label{fig:shuffle-pathological}
\end{figure*}

Figure~\ref{fig:shuffle-pathological} demonstrates $\Rhatmax$ detecting two
artificially-created pathological conditions: a stuck chain and a shifted chain,
both of which correspond to non-convergence of one of the folds.
To be clear, we do not claim that this $\Rhatmax$ benchmark is foolproof
or even that it will detect most convergence issues, but it did perform well
in our examples when the bulk of folds had mixed (Section~\ref{sec:examples}).

\section{Illustrative examples\label{sec:examples}}

In this section we present three additional applied examples
of PCV. Each example uses PCV to select between two candidate
models for a given application. Code can be
found online at \texttt{https://github.com/kuperov/ParallelCV}. For
each candidate model, we first perform model checking on the full-data model
prior to running CV, where chains are initialized using prior draws.
For all experiments,
we fix the batch size $b=50$, and check that this batch size yields
reasonable ESS estimates compared with window methods \citep{kumar2019arviz}
on the full-data models.

The core procedure we use for our experiments requires only a log joint likelihood
function and log predictive density function compatible with JAX's
primitives, and is therefore amenable to automatic vectorization.
Our examples use the HMC and window adaptation implementations
in Blackjax v1.0 \citep{blackjax2020github},
as well as primitives in TensorFlow Probability (TFP; \cite{Dillon2017}). All
experiments use double-precision (64-bit) arithmetic.
Full-data inference is performed on the CPU while
parallel inference is run on the GPU. CPU and GPU details are noted
in each results table.

\begin{example}[Rat weight]
\label{exa:Rats}This example demonstrates PCV on grouped data.
\citet{gelfand1990} present a model
of the weight of $J=30$ rats, for each of which five weight measurements
are available. The rat weights are modeled as a function of time,
\begin{align}
M_{A}: &  & y_{j,t}|\alpha_{j},\beta_{j}, \sigma_y & \sim\mathcal{N}\left(\alpha_{j}+\beta_{j}t,\sigma^{2}_y\right),
\end{align}
for $j=1,\dots,J;t\in\left\{ 8,15,22,29,36\right\}$, 
where $\alpha_j$ and $\beta_j$ denote random effects per rat.
The model $M_{A}$ random effects and per-rat effects are modeled
hierarchically,
\begin{equation}
\alpha_{j}\,|\,\mu_\alpha,\sigma_\alpha \sim\mathcal{N}\left(\mu_{\alpha},\sigma_{\alpha}^{2}\right),\hfill\beta_{j}\sim\mathcal{N}\left(\mu_{\beta},\sigma_{\beta}^{2}\right)
\end{equation}
with hyper-priors $\mu_{\alpha}\sim\mathcal{N}\left(250,20\right)$,
$\mu_{\beta}\sim\mathcal{N}\left(6,2\right)$, $\sigma_{\alpha}\sim\mathrm{Gamma}\left(25,2\right)$,
and $\sigma_{\beta}\sim\mathrm{Gamma}\left(5,10\right)$. The observation noise prior is
$\sigma_y\sim\mathrm{Gamma}(1,2)$. Prior
parameters were chosen using prior predictive checks.

In this example, we use parallel CV to check whether the random effect
(i.e. rat-specific slope $\beta_{j}$) does a better job of predicting
the weight of a new rat, than if a common $\beta$ had been used.
The CV scheme leaves a rat out for each fold, for a total of $K=30$
folds. The alternative model is
\begin{equation}
M_{B}:y_{j,t}|\alpha_{j},\beta,\sigma^{2}\sim\mathcal{N}\left(\alpha_{j}+\beta t,\sigma^{2}\right),
\end{equation}
for $j=1,\dots,J;t\in\left\{ 8,15,22,29,36\right\}$,
where the prior $\beta\sim\mathcal{N}\left(6,2\right)$ was chosen
using prior predictive checks \citep{Gelman2014}.

\begin{figure*}
\begin{centering}
\includegraphics[width=1\textwidth]{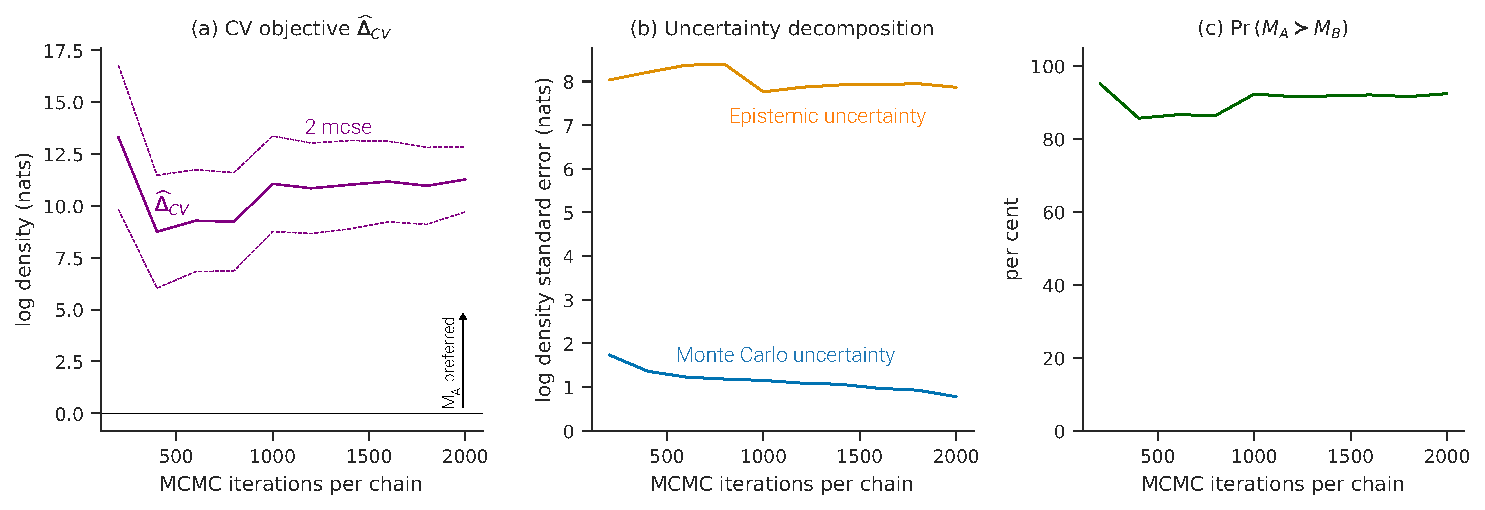}
\par\end{centering}
\caption{Progressive estimates for groupwise cross-validation for Example~\ref{exa:Rats}
(rat weight). The model selection statistic $\widehat{\Delta}$ shown in Panel~(a)
is clearly positive, favoring model $M_A$. Panel~(b) shows that MC uncertainty
is a very small component of the total. Panel~(c) shows that the probability
of $M_{A}$ predicting better than $M_{B}$ stabilizes after about 1,000 iterations.}
\label{fig:ex-rat}
\end{figure*}

Figure~\ref{fig:ex-rat} shows that the PCV results have stabilized by
1,000 iterations, and $\Pr\left(M_{A}\succ M_B\right)\approx90\%$.
On an NVIDIA T4 GPU, PCV with 480 chains
targeting all 60 posteriors took 18 seconds, which included a 10 second
warm-up phase (Table~C2 in Appendix~C). Full-data inference 
took 11 and 15 seconds, respectively, which suggests naive brute
force CV would take about 13 minutes. $\Rhatmax$
plots suggest convergence after around 500 iterations per chain
(Figure~C2, Appendix~C).
In contrast, many chains still exceeded the 1.01 benchmark for
$\Rhat$ even after $2,000$ iterations (Figure~C1, Appendix~C).
\end{example}

\begin{example}[Home radon]
\label{exa:Radon}Radon is a naturally-occurring radioactive element
that is known to cause lung cancer in patients exposed to sufficiently
high concentrations. \citet{gelman2006data} present a hierarchical
model of radon concentrations in U.S. homes. The data cover $N_D=12,573$
homes in $J=386$ counties. For our purposes we will assume that the goal of the model
is to predict the level of radon in U.S. counties, including those
not in the sample (i.e. out-of-sample county-wise prediction). The
authors model the level of radon $y_{i}$ in the $i$th house as normal,
conditional on a random county effect $\alpha_{j}$ and the floor
of the house $x_{i}$ where the measurement was taken. We will compare
two model formulations: 
\begin{align}
M_{A}: &  & y_{i}\,|\,\alpha,\beta,\sigma_{y}^{2} & \sim\mathcal{N}\left(\alpha_{j\left[i\right]}+\beta x_{i},\sigma_{y}^{2}\right),\\
M_{B}: &  & y_{i}\,|\,\alpha,\beta,\sigma_{y}^{2} & \sim\mathcal{N}\left(\alpha_{j\left[i\right]},\sigma_{y}^{2}\right),
\end{align}
for $i=1,\dots,N_D,$, where $\beta$ is a fixed effect, $\alpha_{j\left[i\right]}$ is the
random effect for the county corresponding to observation $i$, and
$\sigma^{2}$ is a common observation variance. For both models the
county effect is modeled hierarchically,
\begin{equation}
\alpha_{j}\,|\,\mu_{\alpha},\sigma_{\alpha}^{2}\sim\mathcal{N}\left(\mu_{\alpha},\sigma_{\alpha}^{2}\right),
\end{equation}
for $j=1,\dots,J$. 
The remaining priors are chosen to be weakly informative, $\mu_{\alpha}\sim \mathcal{N}\left(0,4\right)$
and $\sigma_{\alpha}^{2}\sim\mathrm{Gamma}\left(6,9\right)$. The
other parameter priors are $\beta\sim \mathcal{N}\left(0,1\right)$ and $\sigma_{y}^{2}\sim\mathrm{Gamma}\left(10,10\right)$.
A non-centered parameterization is used for MCMC inference and the
model is fit by HMC. Prior parameters were chosen using prior predictive
checks.

We will use county-wise PCV to determine
whether the floor measure improves predictive performance.
The estimate $\Pr\left(M_{A}\succ M_B\right)\approx100\%$ stabilizes quickly
(Figure~\ref{fig:ex-radon}).

The parallel inference procedure takes a total of 92 seconds
to draw 2,000 parallel
MCMC iterations plus 2,000 warmup iterations across a total of
3,088 chains targeting 772 posteriors. The 386 fold
posteriors are sampled consecutively for each model.
This compares with 45 and 35 seconds for the full-data models (see Table~C3
in Appendix~C for details). At 35 seconds per
fold, a naive implementation of brute force CV across all would have
taken 7.5 hours to run.

\begin{figure*}
\begin{centering}
\includegraphics[width=1\textwidth]{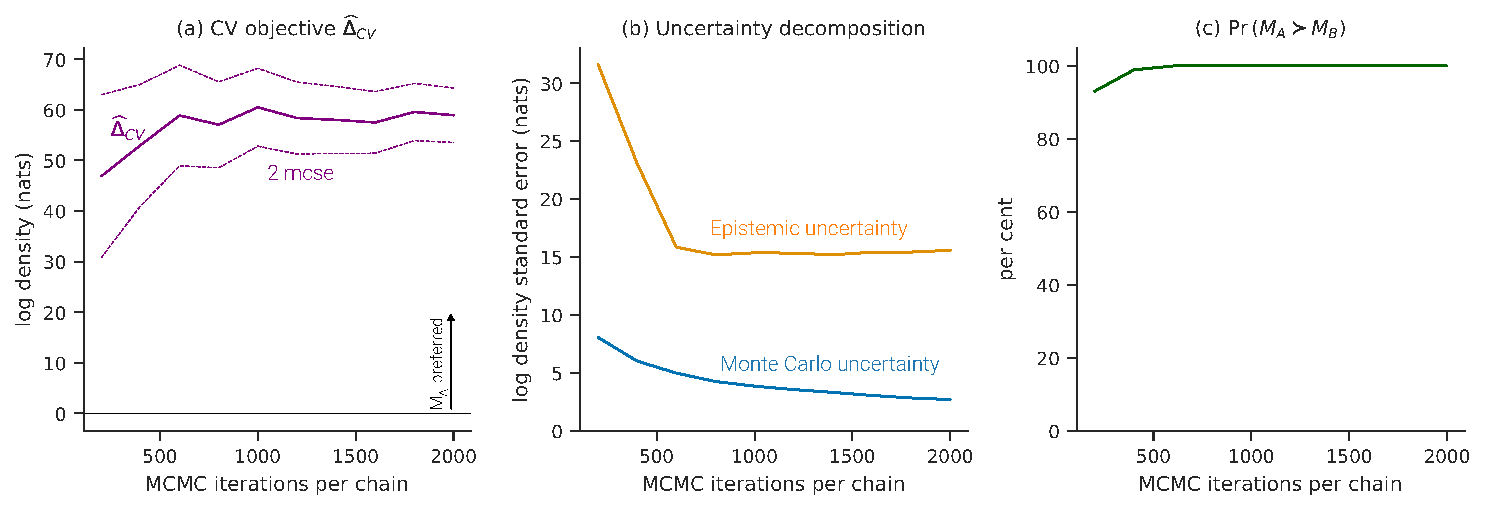}
\par\end{centering}
\caption{Progressive estimates for groupwise cross-validation for Example~\ref{exa:Radon}
(home radon). Model $M_A$ is preferred. The model selection
statistic (a) is clearly positive and the probability that $M_{A}$
predicts better than $M_{B}$ stabilizes within a few hundred iterations.}
\label{fig:ex-radon}
\end{figure*}
\end{example}

\begin{example}[Air passenger traffic to Australia]
\label{ex:flights}Australia is an island nation, for which almost
all migration is by air travel. Models of passenger arrivals and departures
are useful for estimating airport service requirements, the health
of the tourist sector, and economic growth resulting from immigration.

We compare two simple models of monthly international air arrivals
to all Australian airports in the period 1985-2019, using data provided
by the Australian Bureau of Transport and Infrastructure Research
Economic (BITRE). The data are seasonal and nonstationary (Figure~C5,
Appendix~C), so we model month-on-month ($M_{A}$)
and year-on-year ($M_{B}$) changes with a seasonal autoregression.
The power spectrum of the month/month growth rates display seasonality
at several frequencies, while annual figures do not, suggesting that
annual seasonality is present (Figure~C6, Appendix~C).

It is therefore natural to model these series using seasonal autoregressions
on the month/month or year/year growth rates. Let $y_{t}\in\mathbb{R}$
denote the growth rate, observed monthly. We model
\begin{equation}
y_{t}=\sum_{i=1}^{p}\rho_{i}y_{t-i}+\beta_{0}+\sum_{j=1}^{q}\beta_{j}d_{j}+\sigma\varepsilon_{t},
\end{equation}
for $t=1,\dots,T,\quad\varepsilon_{t}\overset{\mathrm{iid}}{\sim}\mathcal{N}\left(0,1\right).$.
The noise standard deviation prior is $\sigma\sim\mathcal{N}^{+}\left(0,1\right)$.
For AR effects we impose the prior $\left(2\rho_{i}-1\right)\sim\mathrm{Beta}\left(5,5\right)$
and for the constant and seasonal effects $\beta_{j}\sim \mathcal{N}\left(0,1\right)$.

Parallel CV results stabilize after a few hundred iterations per chain
(Figure~C4, Appendix~C). PCV took a total of 6.1 seconds to
draw to draw 500 iterations per chain (Table~C4, Appendix~C).
Naively running all 790 models in succession would have taken about
4.4 hours.

\begin{figure*}
\begin{centering}
\includegraphics[width=1\textwidth]{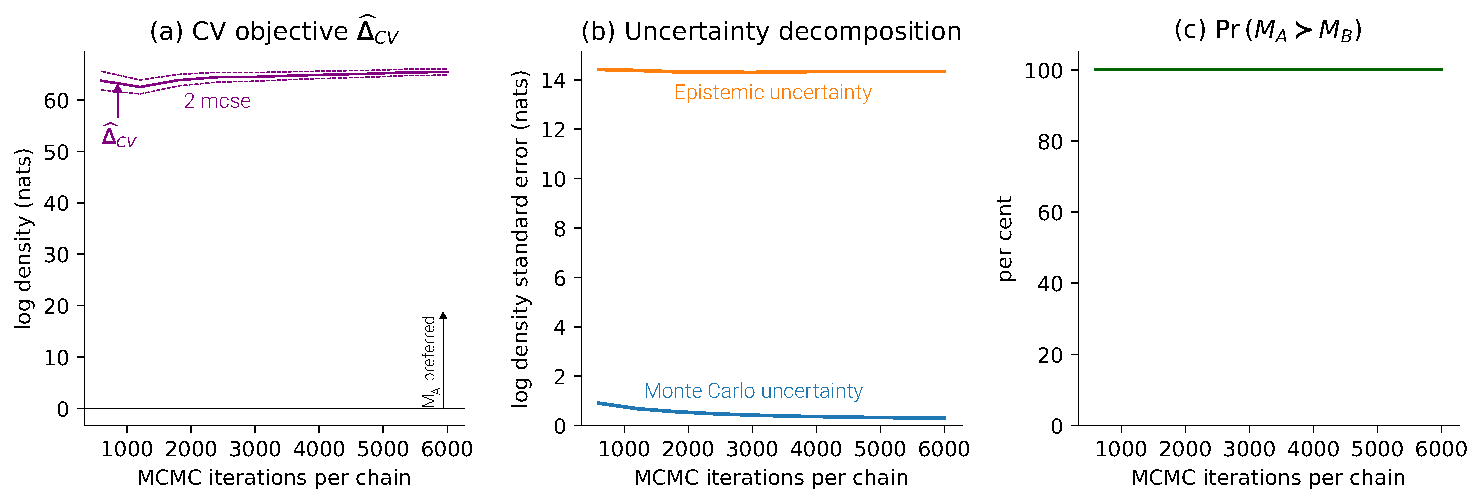}
\par\end{centering}
\caption{Progressive estimates for groupwise cross-validation for Example~\ref{ex:flights},
the model of passenger arrivals. The model selection statistic (a)
is clearly positive and the probability of $M_{A}$ predicting better
than $M_{B}$ stabilizes within a few hundred iterations.}
\label{fig:ex-flights}
\end{figure*}
\end{example}

\section{Discussion}

We have demonstrated a practical workflow for conducting fast, general brute force CV in parallel on modern computing accelerator hardware. We have also
contributed methods
for implementing and checking the resulting output. 

Our proposed workflow is a natural extension of standard Bayesian inference workflows
for MCMC-based inference \citep{Gelman2020}, extended to include
initialization of parallel MCMC chains and joint convergence assessment
for the overall CV objective.

The use of parallel hardware enables
significantly faster CV procedures (in wall clock time),
and on a practical level represents
a sharp improvement in both flexibility and speed over existing CPU-based
approaches. In contrast to approximate CV methods, our approach reflects
a transition from computing environments that are predominantly compute-bound
(where storage and bandwidth are not practically constrained), to a new era with
fewer constraints on computing power but where memory and bandwidth are more limited. Efficient use of parallel hardware can in some cases reduce
the energy required and associated carbon release for compute-heavy tasks,
such as simulation studies involving repeated applications of CV.

Our proposed diagnostic criteria provide the analyst with tools for assessing
convergence across a large number of estimated posteriors, alleviating
the need to examine each posterior individually. Further efficiencies
could be gained by developing formal stopping rules for halting inference
when chains have mixed and the desired accuracy has been attained, although
stopping rules should be applied with care as they can also increase bias
\citep[e.g.,][]{Jones2006,Cowles1999}. We leave this to future research.

Moreover, online algorithms' frugal memory requirements
carries advantages for several classes of users. Top-end GPUs
can run larger models (and/or more folds simultaneously),
while commodity computers (e.g. laptops with less capable integrated GPUs)
can perform a larger range of useful tasks. Beyond accelerator hardware,
our approach may have benefits on CPU-based architectures too, by exploiting
within-core vector units and possibly improving processor cache performance
because of the tight memory footprint of online samplers.

Two possible extensions to this work could further reduce memory footprint
on accelerators. The adaptive subsampling approach of \citet{pmlr-v108-magnusson20a}
would require that only a subset of folds be estimated before a decision became
clear. In addition, the use of stochastic HMC \citep{chen2014stochastic} would permit
that only a subset of the dataset be loaded on the accelerator at any one time.

Further work could include adaptive methods to focus
computational effort in the areas that would most benefit from further
MCMC draws, for example on fold posteriors with the largest MC variance,
as well as a stopping rule to halt inference when a decision is clear.
Turn-key parallel brute-force CV routines would be a useful natural 
extension to probabilistic programming languages. Parallel CV can also
be done using inference methods other than MCMC, such as
variational inference.

\backmatter

\bmhead{Acknowledgments}

The authors would like to thank Charles Margossian for helpful comments.
AC's work was supported in part by an Australian Government Research
Training Program Scholarship. AV acknowledges the Research Council
of Finland Flagship program: Finnish Center for Artificial Intelligence,
and Academy of Finland project (340721). CF acknowledges financial
support under National Science Foundation Grant SES-1921523.

\bibliography{pcv}

\newpage
\onecolumn
\FloatBarrier

\begin{appendices}

\section{Online parallel CV algorithms\label{sec:streaming}}

Scoring rules that can be represented as functions of ergodic averages
of the chains can also be implemented with a online estimator. Specifically, we require that
we can express
\begin{equation}
    \widehat{S}\left(p\left(\tilde{y}\,|\,y\right),\tilde{y}\right) \equiv \phi\left(\hat{\mu},\hat{\Sigma}\right)\qquad
\end{equation}
where $\phi:\mathbb{R}^{d}\times\mathbb{R}^{d\times d}\to\mathbb{R}$ and
$\xi:\Theta\to\mathbb{R}^{d}$ are functions, and
\begin{equation}
    \hat{\mu}=\frac{1}{S}\sum_{s=1}^S\xi\left(\theta^{(s)}\right)\quad\text{and}\quad
    \hat{\Sigma} = \frac{1}{S-1}\sum_{s=1}^S\left[\xi\left(\theta^{(s)}\right)-\hat{\mu}\right]\left[\xi\left(\theta^{(s)}\right)-\hat{\mu}\right]^\top.
\end{equation}
Scoring rules of this form include $\mathrm{LogS}$, $\mathrm{DSS}$, and
$\mathrm{HS}$ (Table~\ref{tbl:mapreduce}). Local scoring rules require only
$\hat{\mu}$.

More general scoring rules, including the popular continuous ranked
probability score (CRPS; \citealp{Gneiting2007}) cannot easily be implemented with
online algorithms as they typically require a full set of draws to
be saved to the accelerator during inference for post processing.
In addition, a need for constant memory precludes the use of many popular
diagnostics, such as trace plots.

\subsection{Online variance estimators}

Estimating $\widehat{\eta}$ and MCMC diagnostics requires estimates
for means and variances of sequences of values, computed without
reference the full history of draws produced during inference.
Computing the mean and variance of a sequence $x_1,\dots,x_N$
without retaining individual values can be achieved using the method
of \citet{Welford1962}.

Consider first the univariate case. We sequentially accumulate the sum
$x_1+x_2+\dots$ in the scalar variable $A_x$ and sum of squares $x_1^2+x_2^2+\dots$ in $A_{x^2}$.
Then the mean $\bar{x}=A_x/N$ and
\begin{equation}
    \var(x) = \frac{1}{N-1}\sum_{i=1}^N\left(x_i - \bar{x}\right)^2
    = \frac{1}{N-1}\left(A_{x^2} - \frac{A_x^2}{N}\right).
\end{equation}

In the vector case, we accumulate the sum in $A_x$ and sum of outer squares in $A_{xx^\top}$.
We will also `center' these estimates as follows.
Let $C\in\mathbb{R}^{d}$ be an estimate of the mean $\bar{x}$, where $d=\mathrm{dim}(\bar{x})$.
We will accumulate the sequence of statistics $A_x^{(i)}$ and
$A_{xx^\top}^{({i})}$ by each iteration evaluating the recursions
\begin{equation}
A_x^{(i)}=A_x^{(i-1)}+\left(x_{i}-C\right),
\qquad
A_{xx^\top}^{(i)}=A_{xx^\top}^{(i-1)}+\left(x_{i}-C\right)\left(x_{i}-C\right)^\top,\label{eq:recursions}
\end{equation}
with $A_x^{(0)}=A_{xx^\top}^{(0)}=0$.
The role of the centering constant $C$ is to ensure that the accumulated values
do not grow too large, leading to numerical overflow.
We only need to allocate memory for a single $d$-vector $A_x^{(i)}$ and $d\times d$ matrix
$A_{xx^\top}^{(i)}$. The mean is given by $\bar{x} = A_x^{(N)}/N + C$, and 
the covariance can be computed as
\begin{align}
    \mathrm{cov}\left(x\right)&=\frac{1}{N-1}\sum_{j=1}^{N}\left(x_{j}-C+C-\bar{x}\right)\left(x_{j}-C+C-\bar{x}\right)^\top\\
    &=\frac{1}{N-1}\left[A_{xx^\top}^{(N)} - \frac{1}{N}A_{x}^{(N)}\left(A_{x}^{(N)}\right)^\top\right].
\end{align}
To compute the covariance of batch means of size $a\in\mathbb{N}$, replace
$x_{i}$ with the $j$th batch mean $\bar{x}_{j}$ and evaluate the
recursions \eqref{eq:recursions} every $a$ iterations.

To stabilize these calculations for small values like probability densities, we store values
in logarithms to avoid numerical underflow (e.g. $U_x = \log A_x$). Increments of the form \eqref{eq:recursions} should
use the numerically stable \verb|logsumexp| function, available on most platforms.
The need to take care to ensure numerical stability is especially acute on 
computing accelerators where lower-precision arithmetic (e.g. 32-bit
or even 16-bit floats) is often preferred for increased performance.

\begin{table*}
\begin{centering}
\begin{tabular}{>{\centering}p{0.1\textwidth}>{\centering}p{0.4\textwidth}>{\centering}p{0.4\textwidth}}
\hline 
\multirow{2}{0.1\textwidth}{\textbf{Objective}} & \textbf{Function $\phi$} & \textbf{Function $\xi$}\tabularnewline
 & $\left(\hat{\mu},\hat{\Sigma}\right)\mapsto$ & $\theta^{\left(s\right)}\mapsto$\tabularnewline
\hline 
\noalign{\vskip0.3cm}
LogS & $\log\left(\hat{\mu}\right)$ & $p\left(\tilde{y}\,|\,\theta^{\left(s\right)}\right)$\tabularnewline
HS & $-2\hat{\mu}_{1}+\hat{\mu}_{2}^{2}$ & $\begin{bmatrix}\frac{\partial^{2}\log p\left(\tilde{y}_i\,|\,\theta^{\left(s\right)}\right)}{\partial\tilde{y}_i^{2}}+\left(\frac{\partial\log p\left(\tilde{y}_i\,|\,\theta^{\left(s\right)}\right)}{\partial\tilde{y}_i}\right)^{2}\\
\frac{\partial\log p\left(\tilde{y}_i\,|\,\theta^{\left(s\right)}\right)}{\partial\tilde{y}_i}
\end{bmatrix}$\tabularnewline
DSS & $\begin{array}{c}
-\log\left|\hat{\Sigma}\right|\\
-\left(y-\hat{\mu}\right)^{\top}\hat{\Sigma}^{-1}\left(y-\hat{\mu}\right)
\end{array}$ & indep. draw from $p\left(\tilde{y}\,|\,\theta^{\left(s\right)}\right)$\tabularnewline[0.6cm]
\hline 
\end{tabular}
\par\end{centering}
\caption{Functions for constructing online estimators of three 
positively-oriented scoring rules:
logarithmic score (LogS), Hyvärinen score (HS), and Dawid-Sebastini
score (DSS). The estimated score \textbf{$\phi:\mathbb{R}^{d}\times\mathbb{R}^{d\times d}\to\mathbb{R}$}
is a function of $\hat{\mu}$ and $\hat{\Sigma}$, respectively the
sample mean and variance of $\left\{ \xi\left(\theta^{(s)}\right):\,s=1,\dots,S\right\} $,
where \textbf{$\xi:\Theta\to\mathbb{R}^{d}$} is a function of individual
MCMC parameter draws. For simplicity, this table drops CV
fold indexes from the notation.}
\label{tbl:mapreduce}
\end{table*}

\subsection{Logarithmic score (LogS)}

Algorithm~\ref{alg:sampler} details a online PCV sampler 
for a single model. This sampler can serve as
a building block for (say) conducting pairwise model assessment.
A sample python implementation can be found in Appendix~E2.

The sampler operates on all folds and chains in parallel. Within each chain, the sampler first runs a warmup loop
(see Section~3). In the main inference loop is divided into a nested hierarchy of MCMC batches
and shuffle blocks. The arrays $U_x$ and $U_{x^2}$
accumulate the log sum of predictive densities and corresponding sum of squares
for each fold and chain, respectively. Similarly, $V_x$ and $V_{x^2}$ accumulate the same for
batch means, with one increment per batch loop.
Each MCMC step, the arrays $Y_x$ and $Y_{x^2}$ accumulate the sum of centered log densities and squared
log densities, respectively, which are required to compute $\Rhat$. $Y_x$ and $Y_{x^2}$ are $K\times L\times D$ arrays,
where the third dimension are the $D$ `shuffle blocks' required to compute the $\widehat{R}_\mathrm{max}$ benchmark.

After sampling, the final loop reshuffles $Y_x$ and $Y_{x^2}$ to construct the samples for the $\widehat{R}_\mathrm{max}$
benchmark.

The following functions are referenced in Algorithm~A2. \verb|Rhat|
computes $\widehat{R}$ from Welford accumulators stored in logs.
Note that full-chain accumulators are assumed here, so the $Y$ variables
referenced in Algorithm~A2 must first be summed over the shuffle
block dimension: $Y_{x}^{c}=\sum_{d=1}^{D}Y_{x}^{\left(\cdot,d\right)}$
and $Y_{x^{2}}^{c}=\sum_{d=1}^{D}Y_{x^{2}}^{\left(\cdot,d\right)}$.
\begin{equation}
\mathtt{Rhat}\left(Y_{x}^{c},Y_{x^{2}}^{c}\right) =\sqrt{\frac{N-1}{N}+\frac{B}{NW}}
\end{equation}
where we have defined
\begin{equation}
W =\frac{1}{L\left(N-1\right)}\sum_{\ell=1}^{L}\left[Y_{x^{2}}^{c}-\frac{1}{N}\left(Y_{x}^{c}\right)^{2}\right]
\quad\text{and}\quad
B =\frac{N}{L-1}\sum_{\ell=1}^{L}\left[\frac{Y_{x}^{c\left(\ell\right)}}{N}-\frac{1}{L}\sum_{\ell'=1}^{L}\frac{Y_{x}^{c\left(\ell'\right)}}{N}\right]^{2}
\end{equation}
The functions \texttt{MCSE} and \texttt{ESS} compute the Monte Carlo standard error an effective sample size, respectively:
\begin{equation}
    \mathtt{MCSE}\left(V_{x},V_{x^{2}}\right) =\sqrt{\sigma_{\hat{\eta}}^{2}/LN}
\qquad\text{and}\qquad
\mathtt{ESS}\left(U_{x},U_{x^{2}},V_{x},V_{x^{2}}\right) =LNs_{\hat{\eta}}^{2}/\sigma_{\hat{\eta}}^{2}
\end{equation}
where we have defined
\begin{equation}
\hat\sigma_{\hat{\eta}}^{2} =\sum_{k=1}^{K}\frac{\hat\sigma_{\hat{f}_{k}}^{2}}{\left(\hat{f}_{k}\right)^{2}}
\qquad
\hat s_{\hat{\eta}}^{2} =\sum_{k=1}^{K}\frac{\hat s_{\hat{f}_{k}}^{2}}{\left(\hat{f}_{k}\right)^{2}}
\end{equation}
\begin{equation}
\hat\sigma_{\hat{f}_{k}}^{2} =\frac{1}{NL}\frac{b}{LN/b-1}\sum_{\ell=1}^{L}\left[\exp\left(V_{x^{2}}^{\left(k,\ell\right)}\right)-\frac{1}{N}\exp\left(2V_{x}^{\left(k,\ell\right)}\right)\right]    
\end{equation}
\begin{equation}
\hat s_{\hat{f}_{k}}^{2} =\frac{1}{NL}\frac{1}{LN-1}\sum_{\ell=1}^{L}\left[\exp\left(U_{x^{2}}^{\left(k,\ell\right)}\right)-\frac{1}{N}\exp\left(2U_{x}^{\left(k,\ell\right)}\right)\right]    
\end{equation}
\begin{equation}
\hat{f}_{k} =\frac{1}{NL}\sum_{\ell=1}^{L}\exp\left(U_{x}^{\left(k,\ell\right)}\right)    
\end{equation}

\begin{algorithm*}
\scriptsize
\begin{algorithmic}
\Require Posterior draws $\left\{\theta^{(\mathrm{fd})}_1,\dots,\theta^{(\mathrm{fd})}_{N_\mathrm{fd}}\right\}$, MCMC kernels $\left\{\mathcal{T}_k\left(\theta,\cdot\right)\right\}_{k=1}^K$, log predictive densities $\left\{\log p\left(\tilde{y}_k\,|\,\theta\right)\right\}_{k=1}^K$, folds $K$, chains $L$, warm-up steps $N_{\mathrm{wu}}$,
blocks $D$, batches per block $G$, batch size $b$
\Ensure $\widehat{S}$, $\widehat{R}$, $\widehat{R}_{\mathrm{max}}$, $\widehat{R}_{\mathrm{max\ rep}}$, $\widehat{ESS}$, $\widehat{MSCE}$.
\Initialize {$K\times L$ arrays $V_x, V_{x^2}, U_x,$ and $U_{x^2}$ with initial entries $-\infty$; 
$K\times L\times G$ arrays $Y$ with initial entries $0$; $K\times R$ array $\widehat{R}_{\mathrm{max\ rep}}$; $K\times L\times R$ arrays $Y_x'$, $Y_x''$ with initial entries $0$; and $K$-vectors $C$ and $\widehat{R}$ with initial entries $0$}
\For{$k \in 1,\dots,K$} in parallel:
\Comment{fold loop}
\For{$\ell \in 1,\dots,L$} in parallel:
\Comment{chain loop}
\State $\theta^{(k,\ell)} \gets \text{draw from }\left\{\theta^{(\mathrm{fd})}_1,\dots,\theta^{(\mathrm{fd})}_{N_\mathrm{fd}}\right\}$
\For{$n \in 1,\dots, N_{\mathrm{wu}}$} sequentially
\Comment{warm-up sampling loop}
    \State $\theta^{(k,\ell)} \gets \text{draw from } \mathcal{T}_k\left(\theta^{(k,\ell)},\cdot\right)$
    \State $C^{(k)} \gets C^{(k)} + \left[\log p\left(\tilde{y}_k\,|\,\theta^{(k,\ell)}\right)\right]/(LN_{\mathrm{wu}})$
\EndFor
\For{$d \in 1,\dots, D$} sequentially
\Comment{block loop}
      \For{$n \in 1,\dots,G$} sequentially
    \Comment{batch loop}
    \State ${Z_{x}}^{(k,\ell)} \gets -\infty$
    \For{$h \in 1,\dots, b$} sequentially
        \Comment{sampling loop}
        \State $\theta^{(k,\ell)} \gets \text{draw from } \mathcal{T}_k\left(\theta^{(k,\ell)},\cdot\right)$
        \State $\hat{s}^{(k,\ell)} \gets \log p\left(\tilde{y}_k\,|\,\theta^{(k,\ell)}\right)$
        \State $Z_x^{(k,\ell)} \gets \log\left[\exp Z_x^{(k,\ell)} + \exp \hat{s}^{(k,\ell)}\right]$
        \State $U_{x^2}^{(k,\ell)} \gets \log \left[ \exp U_{x^2}^{(k,\ell)} + \exp \left(2\hat{s}^{(k,\ell)}\right) \right]$
        \State $Y_x^{(k,\ell,d)} \gets Y_x^{(k,\ell,d)} + \hat{s}^{(k,\ell)} - C^{(k)}$
        \State $Y_{x^2}^{(k,\ell,d)} \gets Y_{x^2}^{(k,\ell,d)} + \left[\hat{s}^{(k,\ell)} - C^{(k)}\right]^2$
        \EndFor
    \State $U_x^{(k,\ell)} \gets \log \left[ \exp U_x^{(k,\ell)} + \exp Z_x^{(k,\ell)} \right] $
    \State $V_x^{(k,\ell)} \gets \log \left[ \exp V_{x}^{(k,\ell)} + \exp \left(Z_x^{(k,\ell)} - \log H\right) \right]$
    \State $V_{x^2}^{(k,\ell)} \gets \log \left[ \exp V_{x^2}^{(k,\ell)} + \exp \left(2Z_x^{(k,\ell)} - 2\log H\right)\right]$
    \EndFor
\EndFor
\EndFor
    \For{$r \in 1,\dots,R$} in parallel:
        \Comment{shuffle draw loop}
        \For{$\ell \gets 1,\dots,L$} in parallel:
        \Comment{chain loop}
            \For{$d \gets 1,\dots,D$} in parallel:
            \Comment{block loop}
                \State $\ell' \gets \text{draw from \{1,\dots,L\} uniformly with replacement}$
                \State ${Y'}_{x}^{(k,\ell,r)} \gets {Y'}_{x}^{(k,\ell,r)} + {Y}_{x}^{(k,\ell',d)}$
                \State ${Y'}_{x^2}^{(k,\ell,r)} \gets {Y'}_{x^2}^{(k,\ell,r)} + {Y}_{x^2}^{(k,\ell',d)}$
            \EndFor
        \EndFor
        \State $\widehat{R}_{\mathrm{max\ rep}}^{(k,r)} \gets \mathtt{Rhat}\left({Y'}_x^{(k,\cdot,r)}, {Y'}_{x^2}^{(k,\cdot,r)}\right)$
    \EndFor
\State $\widehat{R}_k \gets \mathtt{Rhat}\left(\sum_{d=1}^D Y_x^{(k,\cdot,d)}, \sum_{d=1}^D Y_{x^2}^{(k,\cdot,d)}\right)$
\EndFor
\State $\widehat{R}_{\mathrm{max}} \gets \max_k \widehat{R}_k$
\State $\widehat{S} \gets \sum_{k=1}^K\left\{\log\sum_{\ell=1}^L \exp{U_x^{(k,\ell)}} - \log(LDGb)\right\} $
\State $\widehat{ESS} \gets \mathtt{ESS}\left(U_x, U_{x^2}, V_x, V_{x^2}\right)$
\State $\widehat{MCSE} \gets \mathtt{MCSE}\left(V_x, V_{x^2}\right)$
\end{algorithmic}
\caption{Single-model online parallel CV sampler for LogS. Superscripts index array elements; an index of "$\cdot$" locates the vector ranged by that coordinate. By convention $\exp\left(-\infty\right) := 0$. See Section~A2 for function definitions.}
\label{alg:sampler}
\end{algorithm*}

\begin{algorithm*}
\begin{algorithmic}
\Require Full-data posterior draws $\left\{\theta^{(\mathrm{fd})}_1,\dots,\theta^{(\mathrm{fd})}_{N_\mathrm{fd}}\right\}$, per-fold MCMC kernels $\left\{\mathcal{T}_k\left(\theta,\cdot\right)\right\}_{k=1}^K$, per-fold log predictive functions $\left\{\log p\left(\tilde{y}\,|\,\theta^{(k,\ell)}\right)\right\}_{k=1}^K$, fold count $K$, chains per fold $L$, warm-up steps $N_{\mathrm{wu}}$,
samples per chain $N$
\Ensure HS estimate, $\widehat{HS}$.
\Initialize {$K$-vector $C$ and $K \times L$ arrays $Y_{1}$ and $Y_{2}$ with initial entries $0$}
\For{$k \in 1,\dots,K$} in parallel:
\Comment{fold loop}
\For{$\ell \in 1,\dots,L$} in parallel:
\Comment{chain loop}
\State $\theta^{(k,\ell)} \gets \text{draw from }\left\{\theta^{(\mathrm{fd})}_1,\dots,\theta^{(\mathrm{fd})}_{N_\mathrm{fd}}\right\}$
\For{$n \in 1,\dots, N_{\mathrm{wu}}$} sequentially
\Comment{warm-up sampling loop}
    \State $\theta^{(k,\ell)} \gets \text{draw from } \mathcal{T}_k\left(\theta^{(k,\ell)},\cdot\right)$
    \State $C_{1}^{(k)} \gets C_{1}^{(k)} + \frac{1}{LN_{wu}}\sum_{i\in\mathsf{test}_k}\left[\frac{\partial^{2}\log p\left(\tilde{y}_i\,|\,\theta^{(k,\ell)}\right)}{\partial\tilde{y}_i^{2}}+\left(\frac{\partial\log p\left(\tilde{y}_i\,|\,\theta^{(k,\ell)}\right)}{\partial\tilde{y}_i}\right)^{2}\right]$
    \State $C_{2}^{(k)} \gets C_{2}^{(k)} + \frac{1}{LN_{wu}}\sum_{i\in\mathsf{test}_k}\left[\frac{\partial\log p\left(\tilde{y}_i\,|\,\theta^{(k,\ell)}\right)}{\partial\tilde{y}_i}\right]$
\EndFor
    \For{$i \in 1,\dots, N$} sequentially
        \Comment{sampling loop}
        \State $\theta^{(k,\ell)} \gets \text{draw from } \mathcal{T}_k\left(\theta^{(k,\ell)},\cdot\right)$
        \State $Y_{1}^{(k,\ell)} \gets Y_{1}^{(k,\ell)} + \sum_{i\in\mathsf{test}_k}\left[\frac{\partial^{2}\log p\left(\tilde{y}\,|\,\theta^{(k,\ell)}\right)}{\partial\tilde{y}^{2}}+\left(\frac{\partial\log p\left(\tilde{y}\,|\,\theta^{(k,\ell)}\right)}{\partial\tilde{y}}\right)^{2}\right] - C_{1}^{(k)}$
        \State $Y_{2}^{(k,\ell)} \gets Y_{2}^{(k,\ell)} + \sum_{i\in\mathsf{test}_k}\frac{\partial\log p\left(\tilde{y}\,|\,\theta^{(k,\ell)}\right)}{\partial\tilde{y}} - C_{2}^{(k)}$
        \EndFor
    \EndFor
\State $\widehat{HS}_k \gets -\frac{2}{LN}\left(\sum_{\ell=1}^L Y_{1}^{(k,\ell)}\right) - 2C_{1}^{(k,\ell)} + \left[\frac{1}{LN}\left(\sum_{\ell=1}^L Y_{2}^{(k,\ell)}\right) + C_{2}^{(k,\ell)}\right]^2$
\EndFor
\State $\widehat{HS} \gets \sum_{k=1}^K \widehat{HS}_k$
\end{algorithmic}

\caption{Basic online parallel CV sampler for HS. Superscripts index array elements.}
\label{alg:sampler-hs}
\end{algorithm*}

\subsection{Hyvärinen score (HS)}

For a predictive density $q$, HS is defined as
\begin{equation}
\HS\left(q,x\right)=2\Delta\log q\left(x\right)+\left\Vert \nabla\log q\left(x\right)\right\Vert ^{2},\label{eq:HS}
\end{equation}
for $\nabla$ the gradient and $\Delta$ the Laplacian operator. The
$\HS$ is proper and key local \citep{Dawid2015a}. $\HS$ can be
computed comparatively efficiently because it is homogeneous and does
not depend on the normalizing term in the predictive densities \citep{shao2019bayesian}.

The HS can be estimated using an online estimator. We will decompose HS by fold, 
$\mathrm{HS}\left(y\right)=\sum_{t=1}^{T}\mathrm{HS}\left(y_{t}\right)$.
\citet{shao2019bayesian} show that, where exchange of differentiation and integration are justified
and observations in the test set are conditionally independent, the $y_{i_k}$ contribution to
$\mathrm{HS}\left(y_{i}\right)$ is given by
\begin{align}
    2\mathbb{E_\vartheta}\left[\frac{\partial^{2}\log p\left(y_{i_k}\,|\,\vartheta\right)}{\partial y_{i_{k}}^{2}}+\left(\frac{\partial\log p\left(y_{i_k}\,|\,\vartheta\right)}{\partial y_{i_{k}}}\right)^{2}\right]
    -\left(\mathbb{E_\vartheta}\left[\frac{\partial\log p\left(y_{i_k}\,|\,\vartheta\right)}{\partial y_{i_{k}}}\right]\right)^{2},
\end{align}
where $\vartheta\sim p(\theta\,|\,y_{-t})$. In our examples, we have a posterior
sample $\theta^{\left(s\right)}\sim p\left(\theta\,|\,y\right)$,
for $s=1,\dots,S$, so we can estimate these expectations using averages.

\begin{algorithm*}
\begin{algorithmic}
\Require Full-data posterior draws $\left\{\theta^{(\mathrm{fd})}_1,\dots,\theta^{(\mathrm{fd})}_{N_\mathrm{fd}}\right\}$, per-fold MCMC kernels $\left\{\mathcal{T}_k\left(\theta,\cdot\right)\right\}_{k=1}^K$, functions to draw from fold predictives $\left\{p\left(\tilde{y}_{\mathsf{test}_k}\,|\,\theta^{(k,\ell)}\right)\right\}_{k=1}^K$, fold count $K$, chains per fold $L$, warm-up steps $N_{\mathrm{wu}}$,
samples per chain $N$
\Ensure DSS estimate, $\widehat{DSS}$.
\Initialize {$K$-vector $C$ and $K \times L$ arrays $Y_{x}$ and $Y_{xx^\top}$ with initial entries $0$}
\For{$k \in 1,\dots,K$} in parallel:
\Comment{fold loop}
\For{$\ell \in 1,\dots,L$} in parallel:
\Comment{chain loop}
\State $\theta^{(k,\ell)} \gets \text{draw from }\left\{\theta^{(\mathrm{fd})}_1,\dots,\theta^{(\mathrm{fd})}_{N_\mathrm{fd}}\right\}$
\For{$n \in 1,\dots, N_{\mathrm{wu}}$} sequentially
\Comment{warm-up sampling loop}
    \State $\theta^{(k,\ell)} \gets \text{draw from } \mathcal{T}_k\left(\theta^{(k,\ell)},\cdot\right)$
    \State $\tilde{y}_{\mathsf{test}_k}^{(k,\ell)} \gets \text{draw from } p\left(\tilde{y}_{\mathsf{test}_k}\,|\,\theta^{(k,\ell)}\right)$
    \State $C_{x}^{(k)} \gets C_{x}^{(k)} + \tilde{y}_{\mathsf{test}_k}^{(k,\ell)}/\left(LN_{wu}\right)$
    \EndFor
    \For{$i \in 1,\dots, N$} sequentially
        \Comment{sampling loop}
        \State $\theta^{(k,\ell)} \gets \text{draw from } \mathcal{T}_k\left(\theta^{(k,\ell)},\cdot\right)$
        \State $\tilde{y}^{(k,\ell)} \gets \text{draw from } p\left(\tilde{y}_{\mathsf{test}_k}\,|\,\theta^{(k,\ell)}\right)$
        \State $Y_{x}^{(k,\ell)} \gets Y_{x}^{(k,\ell)} + \tilde{y}^{(k,\ell)} - C_{x}^{(k)}$
        \State $Y_{xx^\top}^{(k,\ell)} \gets Y_{xx^\top}^{(k,\ell)} + \left(\tilde{y}^{(k,\ell)} - C_{x}^{(k)}\right)\left(\tilde{y}^{(k,\ell)} - C_{x}^{(k)}\right)^\top$
        \EndFor
    \EndFor
    \State $\hat\mu \gets \frac{1}{LN} \left(\sum_{\ell=1}^L Y_x^{(k,\ell)}\right) + C_{x}^{(k)} $
    \State $\hat\Sigma \gets \frac{1}{LN-1} \left[\left(\sum_{\ell=1}^L Y_{xx^\top}\right) - \frac{1}{LN}\left( \sum_{\ell=1}^L Y_x^{(k,\ell)} \right)\left( \sum_{\ell=1}^L Y_x^{(k,\ell)} \right)^\top\right]$
    \State $\widehat{\DSS}_{k}=-\log\left|\widehat{\Sigma}_{k}\right|
    -\left(y_{\mathsf{test}_{k}}-\widehat{\mu}_{k}\right)^{\top}\widehat{\Sigma}_{k}^{-1}\left(y_{\mathsf{test}_{k}}-\widehat{\mu}_{k}\right)$
\EndFor
\State $\widehat{DSS} \gets \sum_{k=1}^K \widehat{\DSS}_{k}$
\end{algorithmic}
\caption{Basic online parallel CV sampler for DSS. Superscripts index array elements. Elements of
each test set $\tilde{y}_{\mathsf{test}_k}$ are presumed conditionally independent.}
\label{alg:sampler-dss}
\end{algorithm*}

\subsection{Dawid-Sebastini score (DSS)}

The DSS is defined as,
\begin{align} 
\widehat{\DSS}_{M,k}\left(y_{\mathsf{test}_{k}}\right)=-\log\left|\widehat{\Sigma}_{M,k}\right|
-\left(\widehat{\mu}_{M,k}-y_{\mathsf{test}_{k}}\right)^{\top}\widehat{\Sigma}_{M,k}^{-1}\left(\widehat{\mu}_{M,k}-y_{\mathsf{test}_{k}}\right),
\end{align}
where $\widehat{\mu}_{M,k}$ and $\widehat{\Sigma}_{M,k}$ are the
mean and covariance of the predictive distributions for $y_{\mathsf{test}_{k}}$
under fold $k$ for model $M$. To estimate $\widehat{\mu}_{M,k}$
and $\widehat{\Sigma}_{M,k}$ with a online estimator,
accumulate one predictive draw
\begin{equation}
    \tilde{y}_{j'}^{(s)}\sim p\left(\cdot\,|\,y_{-j'},\theta_{M,k}^{\left(s\right)}\right)
\end{equation}
per fold and chain, for each MCMC iteration, and compute sample mean and variance
using the Welford estimator.

\FloatBarrier

\section{Masking for parallel evaluation\label{subsec:masking}}

PCV requires the likelihood and scoring functions for each parallel
fold to incorporate different data subsets, respectively corresponding
to the $\mathsf{train}_{k}$ and $\mathsf{test}_{k}$ sets. One way
GPUs efficiently scale computations is by operating on data in batches,
executing computations as \emph{single instruction multiple data }(SIMD)
programs \citep{Holbrook2021SIMD,warne2022vector}. This requires that each fold's likelihood function be calculated
by the same program code.

\textbf{Parallel fold evaluation.} In our experiments we implemented
CV structures using data masks (arrays of binary indicator variables)
to select the appropriate data subset. For the linear regression example,
the $N$ potential likelihood $\ell_{k}$ and predictive log density
$s_{k}$ given $\beta$ and $\sigma^{2}$ are, for each fold $k=1,\dots,K,$
\begin{align}
\ell_{k}\left(\beta,\sigma^{2}\right) & =\sum_{n=1}^{N}\left[I_{k}(n)\log\mathcal{N}\left(y_{n}\,|\,x_{n}^{\top}\beta,\sigma^{2}\right)\right],
\label{eq:mask-log-prob}
\\
s_{k}\left(\beta,\sigma^{2}\right) & =\sum_{n=1}^{N}\left[I_{k}(n)\log\mathcal{N}\left(y_{n}\,|\,x_{n}^{\top}\beta,\sigma^{2}\right)\right].
\label{eq:mask-log-pred}
\end{align}
In the above $I_{k}(n):=I\left\{n\in\mathsf{test}_k\right\}$ denotes and
indicator function for membership in $\mathsf{test}_k$.
The following example python implementations for a log joint density and log predictive 
show that these expressions can be implemented without branching logic. Importantly,
these functions can be parallelized: it can be evaluated in parallel with multiple
different values for the \verb|fold_id| parameter, even in the case where the folds
are of heterogeneous sizes.

{\scriptsize\begin{verbatim}

import jax.numpy as jnp
from jax.scipy.stats import norm

y = ... # array of length N
X = ... # N*p array of covariates
fold_index = ...  # integer array of length N of fold numbers

def log_joint_density(beta, sigsq, fold_id):
    fold_mask = (fold_index != fold_id)
    log_lik_all = norm.logpdf(y, loc=X @ beta, scale=jnp.sqrt(sigsq))
    log_lik_fold = (log_lik_all * fold_mask).sum()
    return log_prior + log_lik_fold

def log_pred(beta, sigsq, fold_id):
    fold_mask = (fold_index == fold_id)
    log_lik_all = norm.logpdf(y, loc=X @ beta, scale=jnp.sqrt(sigsq))
    return (log_lik_all * fold_mask).sum()

\end{verbatim}}

Admittedly, the masking approach described above does perform unnecessary
calculations. Likelihood contributions are computed for all observations,
including those in the test set. Similarly, predictive score contributions
are computed for observations in the training set. However, on a computing
accelerator where computing power is not practically constrained,
this is a small price to pay for the benefit of parallel computation.

\textbf{Parallel model evaluation.} A similar masking approach can
evaluate different models simultaneously, so long as the
overall structure of the models and parameter are similar enough. The benefit is
that multiple models under consideration can be evaluated in parallel;
otherwise inference needs to be considered sequentially.

Consider a comparison between nested Gaussian linear regression models,
with $J$ potential covariates. Let the binary selection vector $M\in\mathbb{B}^{J}$
define the required model subset, where $M_{j}=1$ denotes inclusion
of the $i$th covariate. Then use the selection vector to zero out
certain data elements. The log likelihood function
$\ell_{k,M}\left(\beta,\sigma^{2}\right)$ has the form
\begin{equation}
\sum_{n=1}^{N}\left[I_k\left(n\right) \log\mathcal{N}\left(y_{n}\,|\,\sum_{j=1}^{J}M_{j}x_{n,j}\beta_{j},\sigma^{2}\right)\right],\label{eq:model-subset}\notag
\end{equation}
and similarly for the score function. Again, this expression
can be evaluated without branching logic.

The masking approach works well for gradient-based inference methods,
such as HMC. Where an element $M_{j}=0$, the corresponding $\beta_{j}$
is simply ignored by the likelihood and predictive. This poses no
problem for inference, since automatic differentiation will correctly
accumulate gradients for $\nabla\ell_{k,M}\left(\beta,\sigma^{2}\right)$,
and HMC can be applied directly. However, leaving parameter elements
unused \emph{can} pose a problem for some automatic hyper-parameter
tuning algorithms that rely on properties of the covariance matrix
of the chains such as MEADS \citep{Hoffman2022}.

\FloatBarrier

\section{Extra figures}\label{sec:figures}


\begin{figure*}
\begin{centering}
\includegraphics[width=1\textwidth]{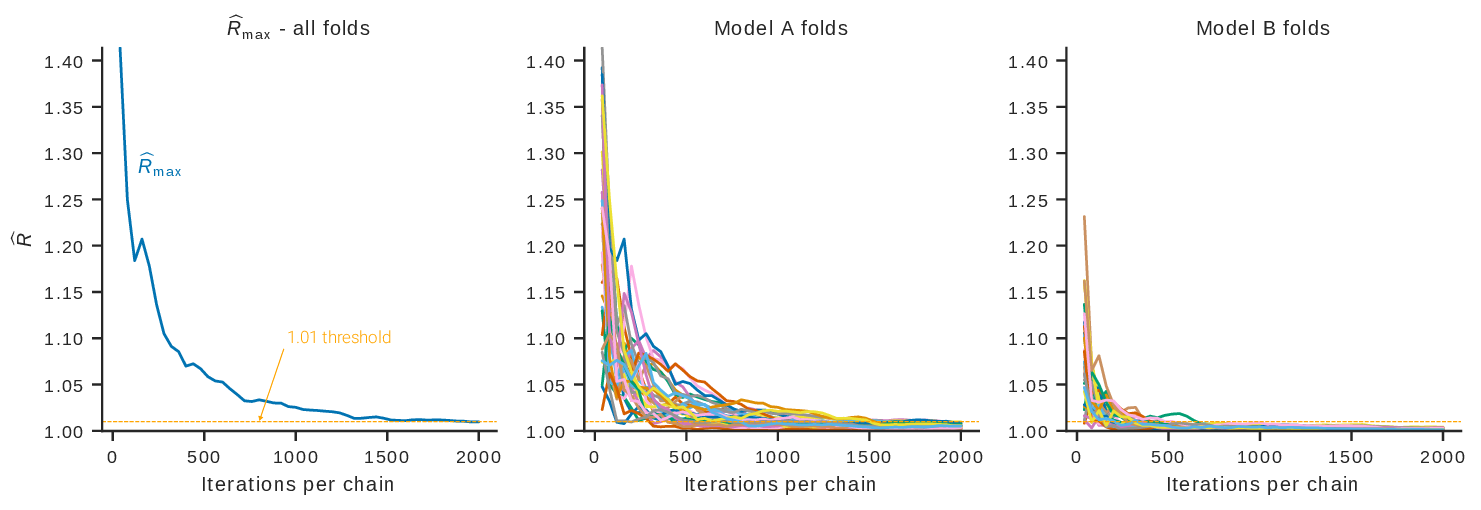}
\par\end{centering}
\caption{Progressive $\Rhat$ statistics for folds of Example~2 (rat weight). At least one fold
$\Rhat$ value (and hence $\Rhatmax$) remains above the 1.01
threshold for at least 1,750 iterations per chain, well beyond the
point where the results have stabilized.}
\label{fig:rat-rhat-chain}
\end{figure*}

\begin{figure*}
\begin{centering}
\includegraphics[width=1\textwidth]{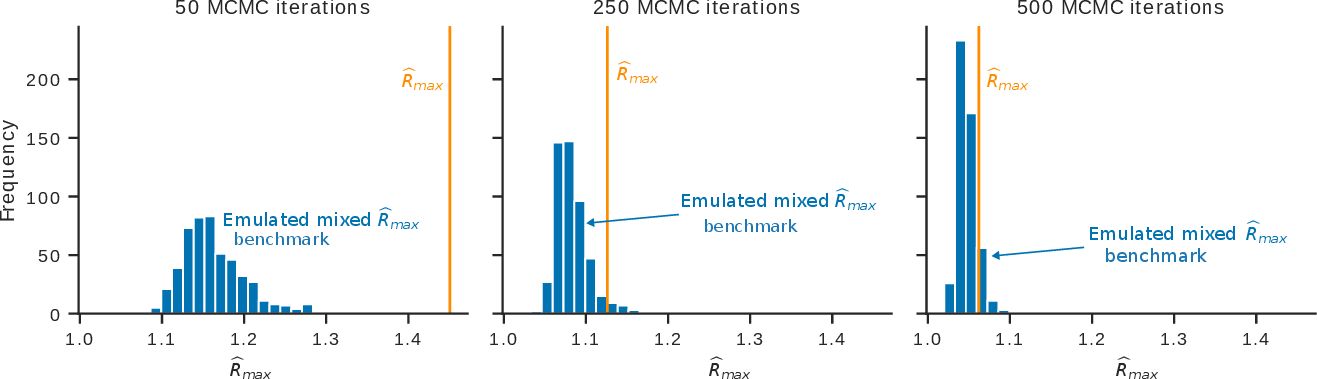}
\par\end{centering}
\caption{$\Rhatmax$ and simulated mixed threshold 
for three different sample sizes in
Example~2 (rat weight). The plots suggest
the chains have not mixed after 50 iterations, but are well
mixed after 500 iterations.}
\label{fig:rat-rhat-eta}
\end{figure*}


\begin{table*}
\begin{centering}
\begin{tabular}{ccr@{\extracolsep{0pt}.}lr@{\extracolsep{0pt}.}lr@{\extracolsep{0pt}.}lr@{\extracolsep{0pt}.}lr@{\extracolsep{0pt}.}lcr@{\extracolsep{0pt}.}lr@{\extracolsep{0pt}.}l}
\hline 
 &  & \multicolumn{4}{c}{Problem size} & \multicolumn{2}{c}{} & \multicolumn{4}{c}{Samples / chain} &  & \multicolumn{4}{c}{Wall time (s)}\tabularnewline
 &  & \multicolumn{2}{c}{chains} & \multicolumn{2}{c}{posteriors} & \multicolumn{2}{c}{} & \multicolumn{2}{c}{warmup} & \multicolumn{2}{c}{sampling} &  & \multicolumn{2}{c}{warmup*} & \multicolumn{2}{c}{sampling}\tabularnewline
\cline{1-1} \cline{3-6} \cline{5-6} \cline{9-12} \cline{11-12} \cline{14-17} \cline{16-17} 
Full-data ($M_{A}$) &  & \multicolumn{2}{c}{8} & \multicolumn{2}{c}{1} & \multicolumn{2}{c}{} & \multicolumn{2}{c}{7,000} & \multicolumn{2}{c}{2,000} &  & 14&7 & 0&5\tabularnewline
Full-data ($M_{B}$) &  & \multicolumn{2}{c}{8} & \multicolumn{2}{c}{1} & \multicolumn{2}{c}{} & \multicolumn{2}{c}{7,000} & \multicolumn{2}{c}{2,000} &  & 10&5 & 0&5\tabularnewline
PCV ($M_{A}$ vs $M_{B}$) &  & \multicolumn{2}{c}{480} & \multicolumn{2}{c}{60} & \multicolumn{2}{c}{} & \multicolumn{2}{c}{1,000} & \multicolumn{2}{c}{500} &  & 10&0 & 7&9\tabularnewline
\hline 
\end{tabular}
\par\end{centering}
\caption{Summary of parallel leave-case-out CV for Example~2 (rat weight).
Inference was performed on a 2.0 GHz Intel
Xeon Cascade Lake-SP CPU with a NVIDA T4 GPU provided by Google Colab.
{*} Includes JAX compilation time.}
\label{tbl:rats}
\end{table*}


\begin{figure*}
\begin{centering}
\includegraphics[width=1\textwidth]{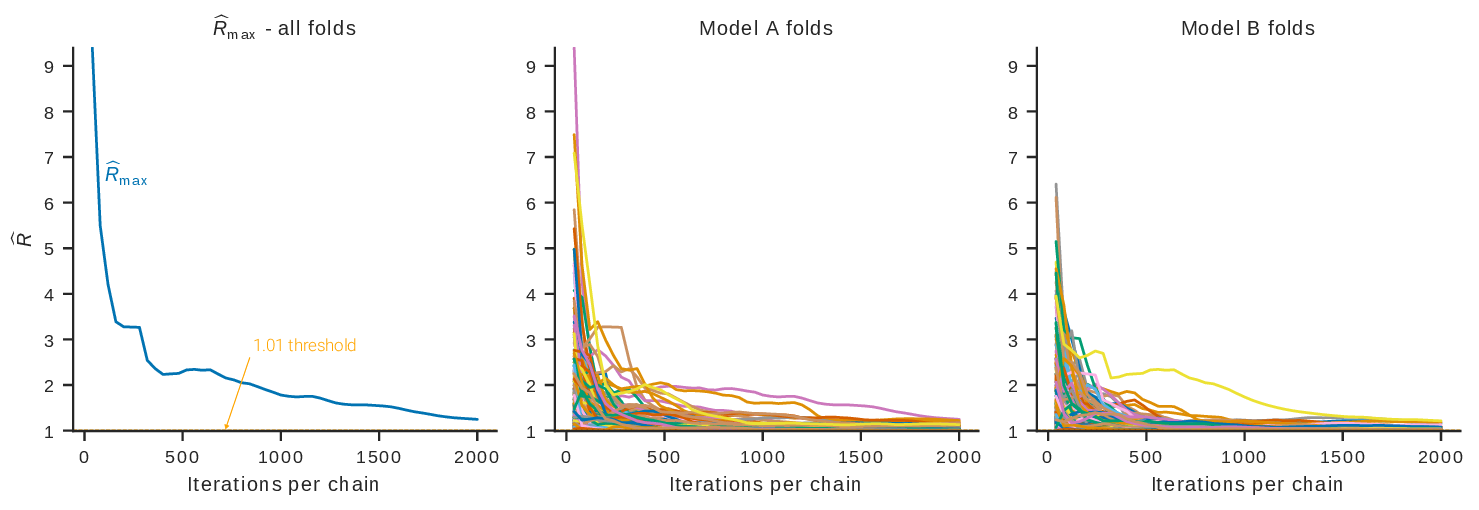}
\par\end{centering}
\caption{Progressive $\Rhat$ statistics for folds of Example~3 (home radon).
$\Rhatmax$ remains above $1.01$ for all 4,000 iterations.}
\end{figure*}

\begin{figure*}
\begin{centering}
\includegraphics[width=1\textwidth]{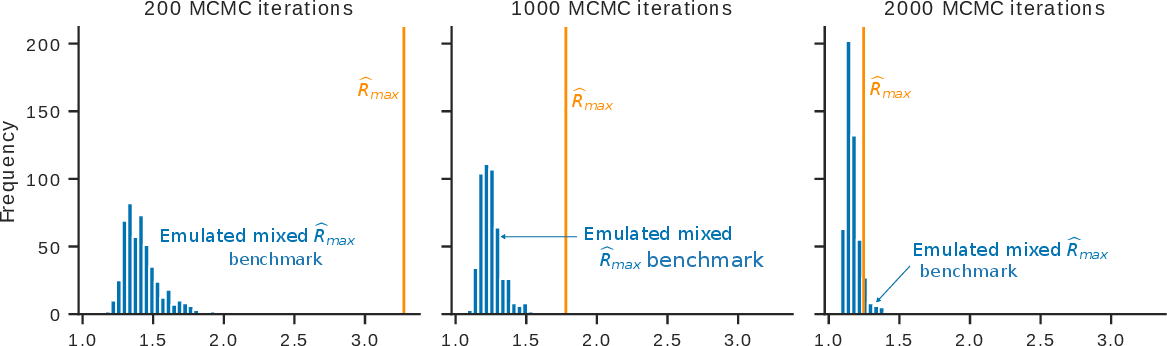}
\par\end{centering}
\caption{$\Rhatmax$ for three different sample sizes in
Example~3 (home radon), under log loss. The benchmark is computed with 5
breaks and 500 draws. The plots suggest
the results have converged by 4,000 iterations. However, notice that
the $\Rhatmax$ value is well above the $1.01$
threshold suggested by \citet{Vehtari2020}.}
\label{fig:radon-conv}
\end{figure*}


\begin{table*}
\begin{centering}
\begin{tabular}{ccr@{\extracolsep{0pt}.}lr@{\extracolsep{0pt}.}lr@{\extracolsep{0pt}.}lr@{\extracolsep{0pt}.}lr@{\extracolsep{0pt}.}lcr@{\extracolsep{0pt}.}lr@{\extracolsep{0pt}.}l}
\hline 
 &  & \multicolumn{4}{c}{Problem size} & \multicolumn{2}{c}{} & \multicolumn{4}{c}{Samples / chain} &  & \multicolumn{4}{c}{Wall time (s)}\tabularnewline
 &  & \multicolumn{2}{c}{chains} & \multicolumn{2}{c}{posteriors} & \multicolumn{2}{c}{} & \multicolumn{2}{c}{warmup} & \multicolumn{2}{c}{sampling} &  & \multicolumn{2}{c}{warmup*} & \multicolumn{2}{c}{sampling}\tabularnewline
\cline{1-1} \cline{3-6} \cline{5-6} \cline{9-12} \cline{11-12} \cline{14-17} \cline{16-17} 
Full-data ($M_{A}$) &  & \multicolumn{2}{c}{4} & \multicolumn{2}{c}{1} & \multicolumn{2}{c}{} & \multicolumn{2}{c}{7,000} & \multicolumn{2}{c}{5,000} &  & 36&8 & 9&4\tabularnewline
Full-data ($M_{B}$) &  & \multicolumn{2}{c}{4} & \multicolumn{2}{c}{1} & \multicolumn{2}{c}{} & \multicolumn{2}{c}{7,000} & \multicolumn{2}{c}{5,000} &  & 28&8 & 7&9\tabularnewline
PCV ($M_{A}$ vs $M_{B}$) &  & \multicolumn{2}{c}{3,088} & \multicolumn{2}{c}{772} & \multicolumn{2}{c}{} & \multicolumn{2}{c}{2,000} & \multicolumn{2}{c}{2,000} &  & 47&1 & 45&2\tabularnewline
\hline 
\end{tabular}
\par\end{centering}
\caption{Summary of parallel CV for Example~3 (home radon). 
Inference was performed on a 2.0 GHz Intel Xeon Cascade Lake-SP CPU
with a NVIDA A100 GPU provided by Google Colab. {*} Warmup includes JAX compilation.}
\label{tbl:radon}
\end{table*}


\begin{figure*}
\begin{centering}
\includegraphics[width=1\textwidth]{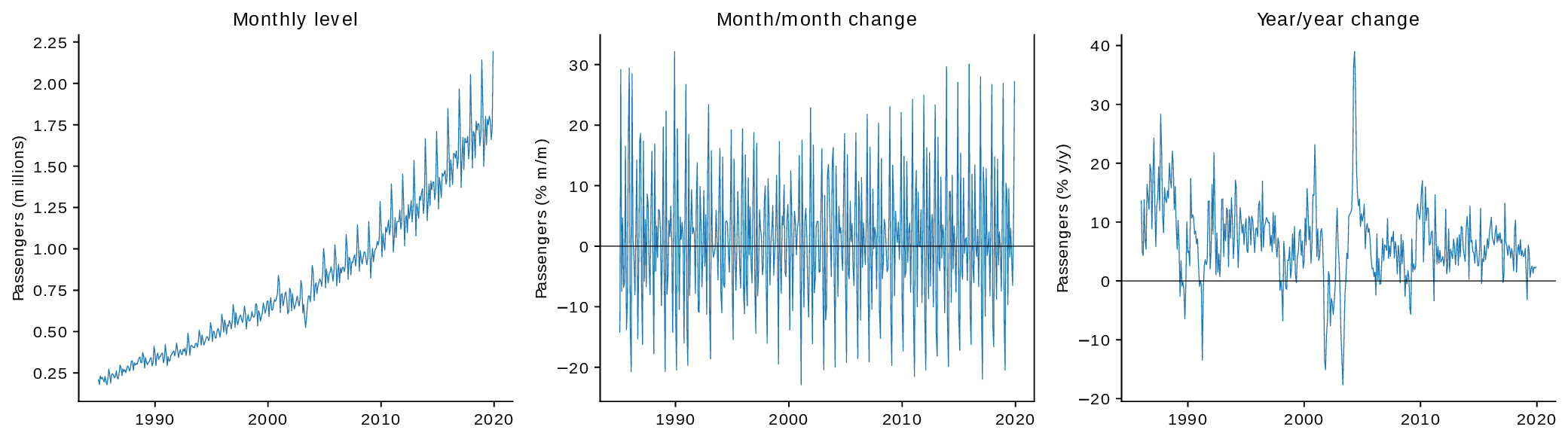}
\caption{International air departures from all Australian airports, 1985-2019.
Source: Australian Bureau of Transport and Infrastructure Research
Economic (BITRE)}
\par\end{centering}
\label{fig:air-arrivals}
\end{figure*}

\begin{figure*}
\begin{centering}
\includegraphics[width=1\textwidth]{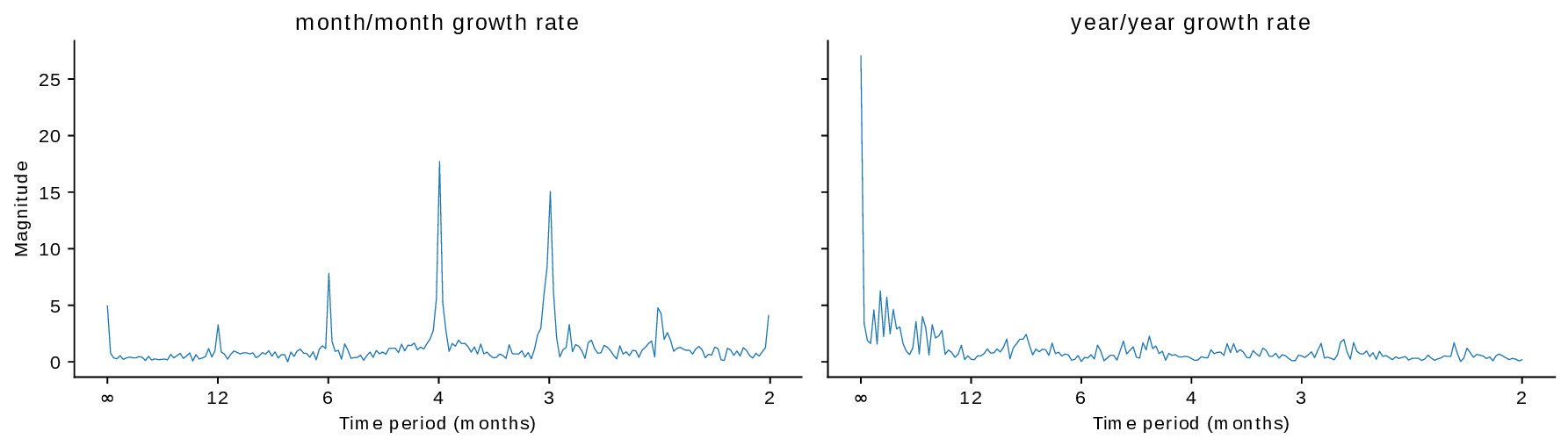}
\caption{Power spectrum of month/month and year/year growth rates for international
air departures from all Australian airports, 1985-2019. Source: Australian
Bureau of Transport and Infrastructure Research Economic (BITRE) and
author's calculations.}
\par\end{centering}
\label{fig:air-spectrum}
\end{figure*}

\begin{figure*}
\begin{centering}
\includegraphics[width=1\textwidth]{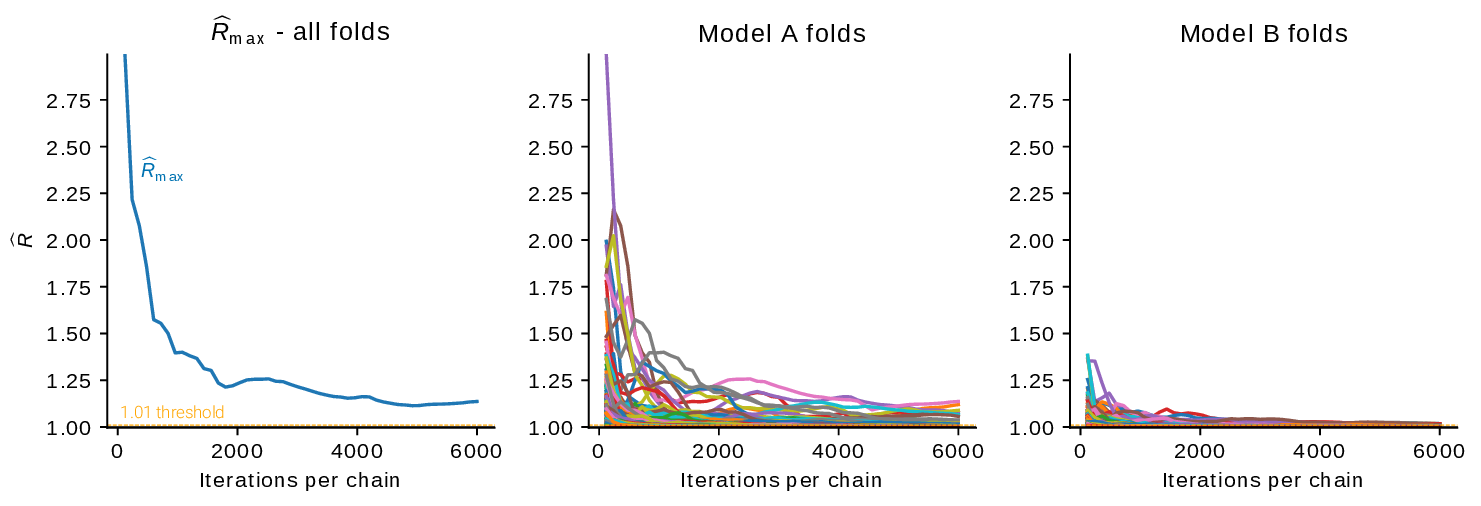}
\par\end{centering}
\caption{Progressive $\Rhat$ statistics for folds of Example~4 (air passengers).
$\Rhatmax$ remains well above the 1.01 threshold for all 2,000 iterations.}
\label{fig:flights-rhat}
\end{figure*}

\begin{figure*}
\begin{centering}
\includegraphics[width=1\textwidth]{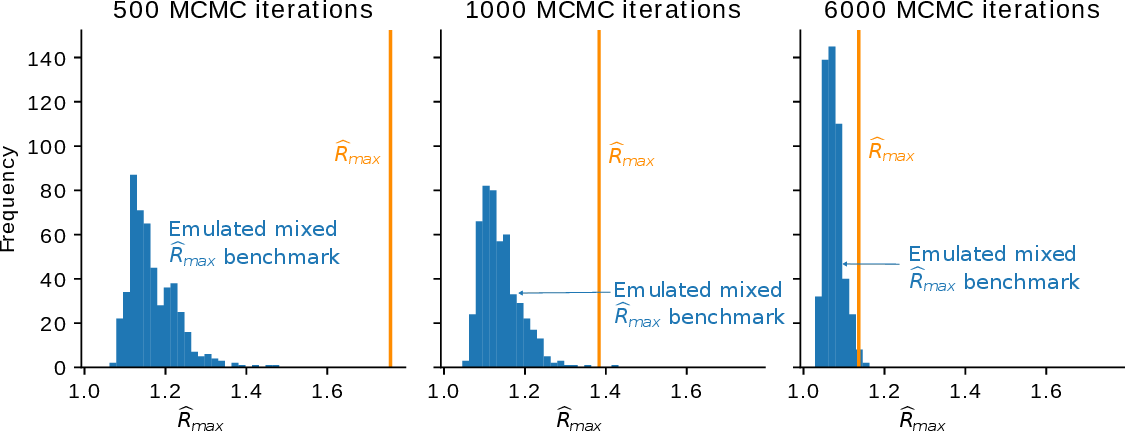}
\par\end{centering}
\caption{$\Rhat_{\widehat{\eta}}$ for three different sample sizes in
Example~4 (air passengers). The plots suggest
the results have converged at 2,000 iterations.}
\label{fig:flights-conv}
\end{figure*}

\begin{table*}
\begin{centering}
\begin{tabular}{ccr@{\extracolsep{0pt}.}lr@{\extracolsep{0pt}.}lr@{\extracolsep{0pt}.}lr@{\extracolsep{0pt}.}lr@{\extracolsep{0pt}.}lcr@{\extracolsep{0pt}.}lr@{\extracolsep{0pt}.}l}
\hline 
 &  & \multicolumn{4}{c}{Problem size} & \multicolumn{2}{c}{} & \multicolumn{4}{c}{Samples / chain} &  & \multicolumn{4}{c}{Wall time (s)}\tabularnewline
 &  & \multicolumn{2}{c}{chains} & \multicolumn{2}{c}{posteriors} & \multicolumn{2}{c}{} & \multicolumn{2}{c}{warmup} & \multicolumn{2}{c}{sampling} &  & \multicolumn{2}{c}{warmup{*}} & \multicolumn{2}{c}{sampling}\tabularnewline
\cline{1-1} \cline{3-6} \cline{5-6} \cline{9-12} \cline{11-12} \cline{14-17} \cline{16-17} 
Full-data ($M_{A}$) &  & \multicolumn{2}{c}{4} & \multicolumn{2}{c}{1} & \multicolumn{2}{c}{} & \multicolumn{2}{c}{10,000} & \multicolumn{2}{c}{2,000} &  & 19&3 & 2&5\tabularnewline
Full-data ($M_{B}$) &  & \multicolumn{2}{c}{4} & \multicolumn{2}{c}{1} & \multicolumn{2}{c}{} & \multicolumn{2}{c}{10,000} & \multicolumn{2}{c}{2,000} &  & 15&4 & 2&0\tabularnewline
PCV ($M_{A}$ vs $M_{B}$) &  & \multicolumn{2}{c}{3,160} & \multicolumn{2}{c}{790} & \multicolumn{2}{c}{} & \multicolumn{2}{c}{6,000} & \multicolumn{2}{c}{6,000} &  & 12&9 & 9&8\tabularnewline
\hline 
\end{tabular}
\par\end{centering}
\caption{Summary of parallel CV for Example~4 (air passengers).
Inference was performed on a 2.0 GHz Intel
Xeon Skylake CPU with a NVIDA T4 GPU provided by Google
Colab. {*}Warmup includes JAX JIT compilation step.}
\label{tbl:flights}
\end{table*}

\FloatBarrier

\section{Example code}\label{sec:excode}

This section presents minimal implementations of parallel CV sampler
for $\LogS$ using python 3.10 and the JAX and BlackJAX libraries. For
simplicity, the code presented here assesses predictive ability from
a single model (it does not perform model selection).
For a complete examples, please see \url{https://github.com/kuperov/ParallelCV}.

\subsection{Non-online parallel sampler}

{\scriptsize\begin{verbatim}

import jax
import jax.numpy as jnp

def pcv_LogS_sampler(key, log_dens_fn, log_pred_fn, init_pos, K, L, N, kparam):
    """Non-online sampler for parallel cross-validation using LogS for a single model.
    
    Generates the predictive draws required to compute the elpd. Invoke this function
    twice to implement Algorithm 1 (once for warmup, once for sampling).

    Args:
        key: JAX PRNG key array
        log_dens_fn: log density function with signature (params, fold_id)
        log_pred_fn: log predictive density function with signature (params, fold_id)
        init_pos: K*L*p pytree of initial positions for each fold and chain
        K: number of folds
        L: number of chains
        N: chain length
        kparam: dictionary of hyperparameters for Blackjax HMC kernel
    
    Returns:
        Tuple: (lpred_draws, E)
    """
    def run_chain(init_pos, chain_key, fold_id, C):  # sample from a single chain
        fold_log_dens_fn = lambda params: log_dens_fn(params, fold_id)
        hmc_kernel = bj.hmc(fold_log_dens_fn, **kparam)

        def mcmc_step(carry_state, _): # a single mcmc step
            key, prev_state, E = carry_state
            step_key, carry_key = jax.random.split(key)
            state, info = hmc_kernel.step(step_key, prev_state)  # one mcmc step
            lpred_draw = log_pred_fn(state.position, fold_id)  # cond. log predictive
            E = E + info.is_divergent
            return (carry_key, state, E), lpred_draw

        init_state = hmc_kernel.init(init_pos)
        return jax.lax.scan(mcmc_step, (chain_key, init_state, 0), None, length=N)

    def run_fold(fold_key, ch_init_pos, fold_id, C): # run L chains for one fold
        sampling_fn = jax.vmap(lambda pos, key: run_chain(pos, key, fold_id, C))
        return sampling_fn(ch_init_pos, jax.random.split(fold_key, L))

    (_, _, E), lpred_draws = \
        jax.vmap(run_fold)(jax.random.split(key, K), init_pos, jnp.arange(K))
    return (lpred_draws, E)

\end{verbatim}}

\subsection{Online parallel sampler}
The function \verb|pcv_LogS_sampler| samples from $K\times L$ posteriors
in parallel and computes the statistics $V_x$, $V_{x^2}$, $U_x$, $U_{x^2}$, $Y_x$, $Y_{x^2}$, which are required to compute 
$\hat{\eta}$, $\widehat{ESS}$, and $\widehat{R}_{\mathrm{max}}$ (Algorithm~A2). The divergence count $E$ requires no further 
computation.

{\scriptsize\begin{verbatim}
    
def pcv_LogS_sampler(key, log_dens_fn, log_pred_fn, init_pos, C_k, L, H, G, D, kparam):
    """Sampler for parallel cross-validation using LogS for a single model.
    
    Generates the statistics required to estimate ESS, MCSE, Rhat_max, and
    the Rhat_max benchmark. Space complexity is O(K*L*D), independent of
    G and H and therefore constant with respect to MCMC chain length.

    Args:
        key: JAX PRNG key array
        log_dens_fn: log density function with signature (params, fold_id)
        log_pred_fn: log predictive density function with signature (params, fold_id)
        init_pos: K*L*p pytree of initial positions for each fold and chain
        C_k: K-array of centering constants per fold
        L: number of chains
        H: MCMC draws per batch
        G: number of batches per block
        D: number of shuffle blocks
        kparam: dictionary of hyperparameters for Blackjax HMC kernel
    
    Returns:
        Tuple: (last_state, Ux, Ux2, Vx, Vx2, Yx, Yx2, E)
    """
    K = C_k.shape[0]

    def run_chain(init_pos, chain_key, fold_id, C):  # sample from a single chain
        fold_log_dens_fn = lambda params: log_dens_fn(params, fold_id)
        hmc_kernel = bj.hmc(fold_log_dens_fn, **kparam)

        def mcmc_step(carry_state, _): # a single mcmc step
            key, prev_state, Zx, Ux2, Yx, Yx2, E = carry_state
            step_key, carry_key = jax.random.split(key)
            params, info = hmc_kernel.step(step_key, prev_state)  # one mcmc step
            lpred = log_pred_fn(params.position, fold_id)  # cond. log predictive
            E = E + info.is_divergent
            Zx = jnp.logaddexp(Zx, lpred)  # increment accumulators
            Ux2 = jnp.logaddexp(Ux2, 2*lpred)
            Yx += lpred - C
            Yx2 += (lpred - C)**2
            return (carry_key, params, Zx, Ux2, Yx, Yx2, E), None

        def batch_step(batch_carry, _): # one batch of H mcmc steps
            key, init_state, Vx, Vx2, Ux, Ux2, Yx, Yx2, E = batch_carry
            init_carry = (key, init_state, -jnp.inf, Ux2)
            (carry_key, state, Zx, Ux2), _ = \
                jax.lax.scan(mcmc_step, init_carry, None, length=H)
            Zx_bar = Zx - jnp.log(H)  # this batch mean
            Vx = jnp.logaddexp(Vx, Zx_bar)  # increment accumulators
            Vx2 = jnp.logaddexp(Vx2, 2*Zx_bar)
            Ux = jnp.logaddexp(Ux, Zx)
            return (carry_key, state, Vx, Vx2, Ux, Ux2, Yx, Yx2, E), None

        def block_step(block_carry, _): # one block of G batches
            init_carry = block_carry + (0, 0,)
            (key, prev_state, Ux, Ux2, Vx, Vx2, Yx, Yx2, E), _ = \
                jax.lax.scan(batch_step, init_carry, None, length=G)
            return (key, prev_state, Ux, Ux2, Vx, Vx2, E), (Yx, Yx2)

        init_state = hmc_kernel.init(init_pos)
        init_carry = (chain_key, init_state, -jnp.inf, -jnp.inf, -jnp.inf, -jnp.inf)
        return jax.lax.scan(block_step, init_carry, None, length=D)

    def run_fold(fold_key, ch_init_pos, fold_id, C): # run L chains for one fold
        sampling_fn = jax.vmap(lambda pos, key: run_chain(pos, key, fold_id, C))
        return sampling_fn(ch_init_pos, jax.random.split(fold_key, L), fold_id)

    (_, last_state, Vx, Vx2, Ux, Ux2, E), (Yx, Yx2) = \
        jax.vmap(run_fold)(jax.random.split(key, K), init_pos, jnp.arange(K), C_k)
    return (last_state, Ux, Ux2, Vx, Vx2, Yx, Yx2, E)
\end{verbatim}}

\end{appendices}

\end{document}